\documentstyle[12pt,epsfig,amsfonts]{article}
\renewcommand{\theequation}{\arabic{section}.\arabic{equation}}

\title{\Huge\bf Interference of outgoing electromagnetic waves\protect
\\generated by two point-like sources}
\author{\bf Yurij Yaremko\\
\normalsize Institute for Condensed Matter Physics\\
\normalsize 1 Svientsitskii St., 79011 Lviv, Ukraine}
\date{August 7, 2003 - February 17, 2004}
\begin{document}
\maketitle
\begin{abstract}
An energy-momentum carried by electromagnetic field produced by two
point-like charged particles is calculated. Integration region considered in
the evaluation of the bound and emitted quantities produced by all points of
world lines up to the end points at which particles' trajectories puncture
an observation hyperplane $y^0=t$. Radiative part of the energy-momentum
contains, apart from usual integrals of Larmor terms, also the sum of work
done by Lorentz forces of point-like charges acting on one another.
Therefore, the combination of wave motions (retarded Li\'enard-Wiechert
solutions) leads to the interaction between the sources.
\end{abstract}

\section{Introduction}
\setcounter{equation}{0}

We consider a closed system of two point electric charges and their
electromagnetic field. A charge $e_a$ produces an electromagnetic
vector potential $A_a^\alpha$ that satisfies the wave equation
\begin{equation} \label{w_e}
\Box A_a^\alpha=-4\pi j_a^\alpha
\end{equation}
together with the Lorentz gauge condition $\partial_\alpha A_a^\alpha=0$.
The vector $j_a^\alpha$ is the charge's current density which is zero
everywhere, except at the particle's position it is infinite. For
concreteness we imagine that the particles are asymptotically free in the
remote past.

The dynamics of electromagnetic field is governed by Maxwell equations with
point-like sources. The action of the field of one source on another is
described by Lorentz force. The evolution of $a$-th particle is
determinated by the relativistic generalization of Newton's second law
where loss of energy due to radiation is taken into account.

The dynamics of the entire system is governed by the action
\begin{equation}\label{S}
S=\sum\limits_{a=1}^2\left(-m_a\int d\tau_a\sqrt{-({\dot
z}_a)^2}+e_a\int d\tau_a A_{a,\mu}{\dot z}_a^\mu\right)
-\frac{1}{16\pi}\int d^4 y f_{\mu\nu}f^{\mu\nu}
\end{equation}
where $f_{\mu\nu}=\sum_a(\partial_\mu A_{a,\nu}-\partial_\nu A_{a,\mu})$.
($a$-th point particle carries electric charge $e_a$ and moves on a
world line $\zeta_a$ described by functions $z_a^\mu(\tau_a)$, in which
$\tau_a$ is an evolution parameter; ${\dot z}_a^\mu:=dz_a^\mu/d\tau_a$.)
Variation on field variables $A_a^\alpha$ yields the Maxwell equations.
Li\'enard-Wiechert fields are the solutions of Maxwell equations with
point-like sources.

Since the field $f_{a,\mu\nu}:=\partial_\mu A_{a,\nu}-\partial_\nu
A_{a,\mu}$ generated by $a$-th source has a singularity on its world line,
demanding that the total action (\ref{S}) be stationary under a variation
$\delta z_a^\mu(\tau_a)$ of the world line does not give sensible motion
equations. To make sense of the retarded field's action on the particle we
should perform the so-called renormalization procedure. It involves
manipulation of the divergent self-energy of a point charge. As usual, the
infinite Coulomb-like term is linked with the "bare" mass $m_a$, so that
the renormalized mass of particle is considered to be finite.

The principle of least action (\ref{S}) is invariant under ten
infinitesimal transformations which constitute the Poincar\'e group.
According to Noether's theorem, these symmetry properties imply
conservation laws, i.e. those quantities that do not change with time. In
his classical paper \cite{Dir}, Dirac used retarded Li\'enard-Wiechert
solution in the law of conservation of the total four-momentum of a
composite (one particle plus field, its own and external) system. It
provides the foundation for his derivation of the radiation-reaction force.
L\'opez and Villarroel \cite{LV} substitute the retarded
Li\'enard-Wiechert field in the angular momentum conserved quantity which
arises from the invariance of the system under space rotations and Lorentz
transformations. The authors arrive at the angular momentum balance
equations which is consistent with the Lorentz-Dirac equation.

\begin{figure}[t]
\begin{center}
\epsfclipon
\epsfig{file=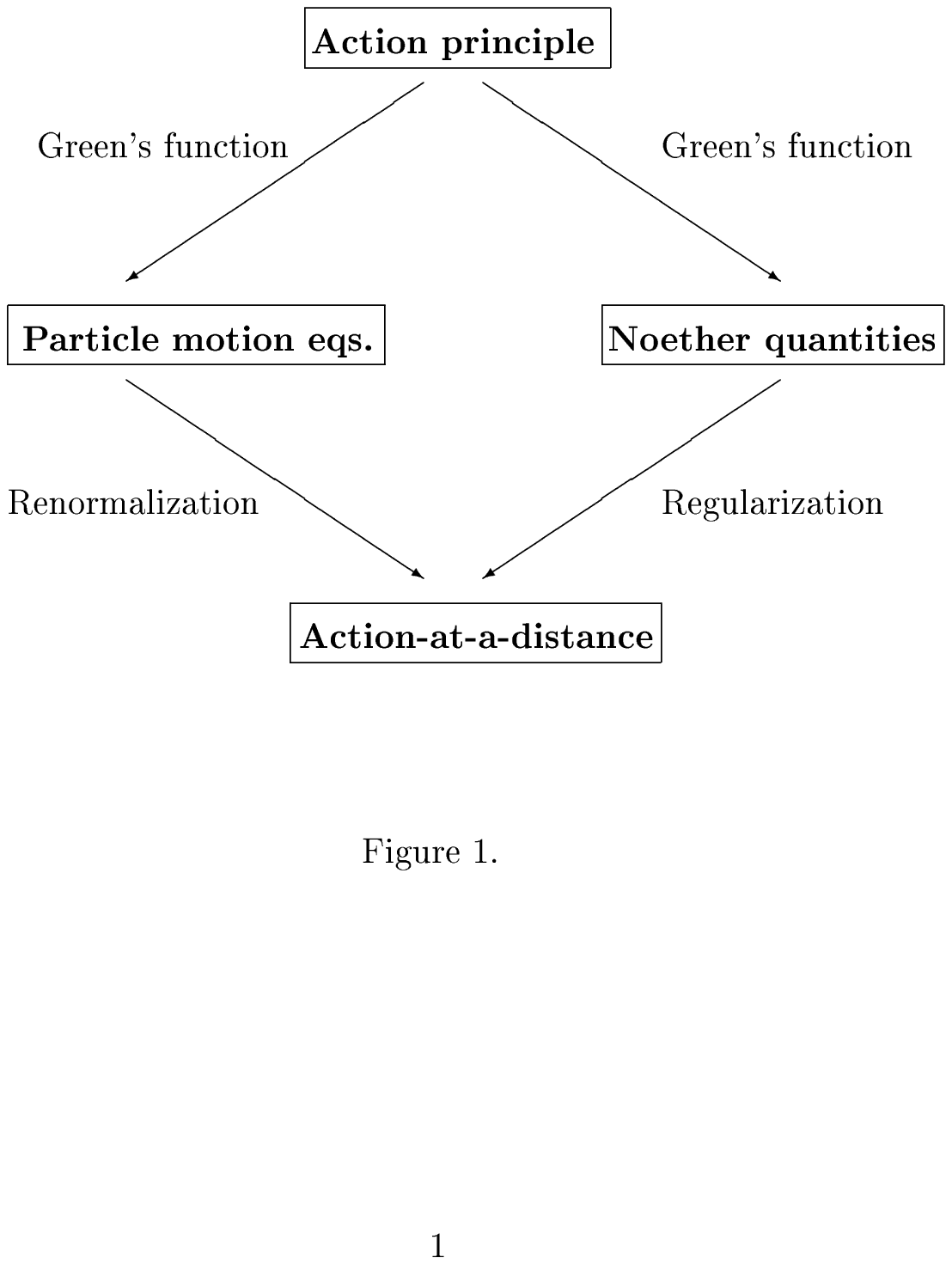,width=10cm}
\end{center}
\caption{\label{regul}
\small The regularization procedure can be performed in two different
ways: (i) one when Green's functions are used in variational equations of
motion; (ii) the other when wave solutions are used in Noether
conservation laws.
}
\end{figure}

\begin{figure}[t]
\begin{center}
\epsfclipon
\epsfig{file=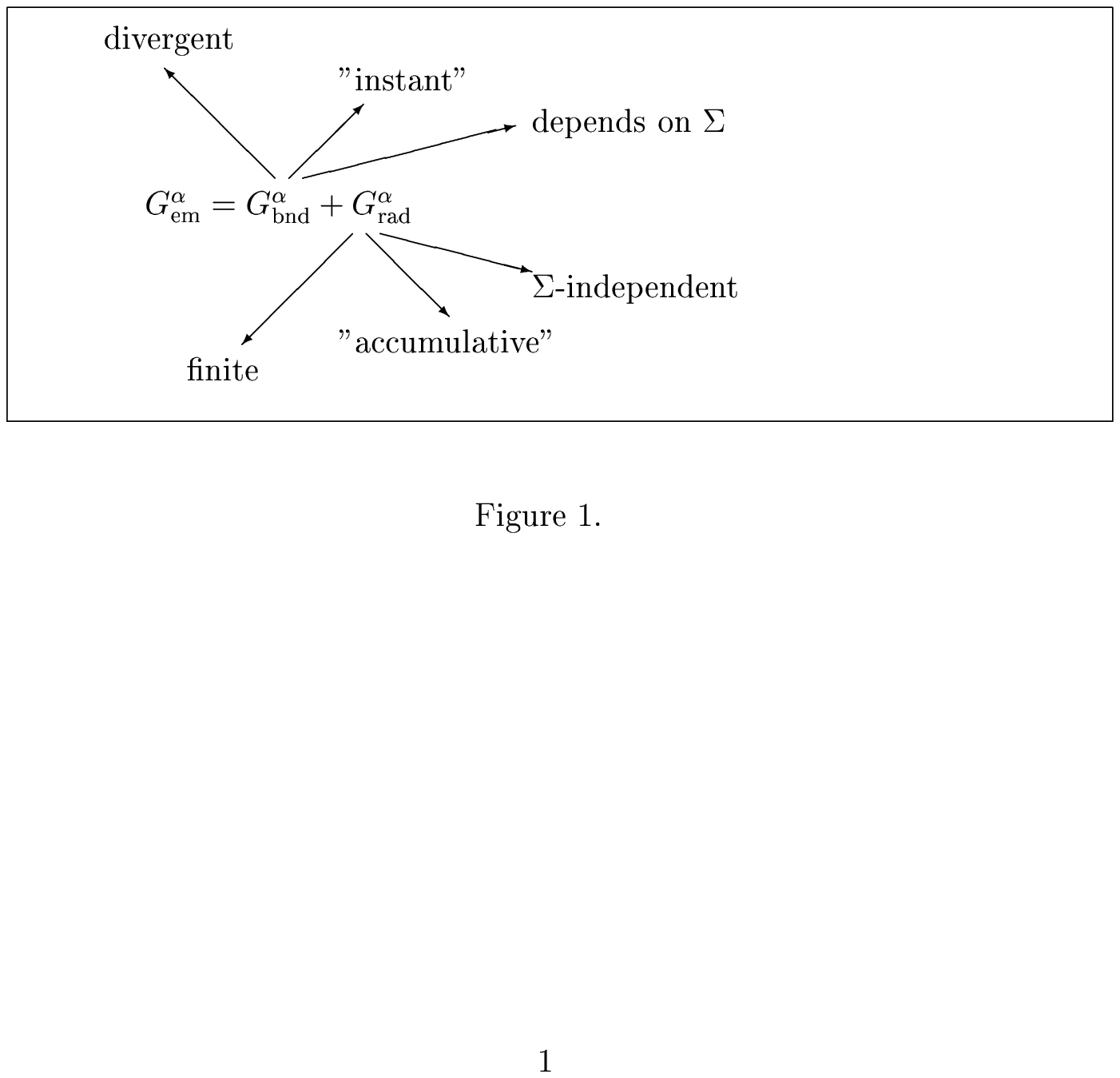,width=8cm}
\end{center}
\caption{\label{Noe}
\small The bound term $G^\alpha_{bnd}$ and the radiative term
$G^\alpha_{rad}$ constitute Noether quantity $G^\alpha_{em}$ carried by
electromagnetic field. The former diverges while the latter is finite.
Bound component depends on instant characteristics of charged particles
while the radiative one is accumulated with time. The form of the bound
term heavily depends on choosing of an integration surface $\Sigma$
while the radiative term does not depend on $\Sigma$.
}
\end{figure}

To find out Noether quantities $G^\alpha_{em}$ carried by
electromagnetic field we integrate the Maxwell stress-energy tensor and
angular momentum tensor density over a space-like three-surface
\cite{Yar1,WFY,RYar,Yar6D}. We obtain terms of two quite different types:
(i) bound, $G^\alpha_{bnd}$, which are permanently "attached" to the
sources and carried along with them; (ii) radiative, $G^\alpha_{rad}$,
which detach themselves from the charges and lead independent existence
(see Fig.\ref{Noe}). Within regularization procedure the bound terms are
coupled with energy-momentum and angular momentum of "bare" sources, so
that already renormalized characteristics $G^\alpha_{part}$ of charged
particles are proclaimed to be finite. Noether quantities which are
properly conserved become:
\begin{equation}
G^\alpha=G^\alpha_{part}+G^\alpha_{rad}.
\end{equation}

Recently \cite{WFY} a frontal collision of two asymptotically free
charges has been considered. We have calculated how much electromagnetic
field momentum and angular momentum flow across hyperplane
$\Sigma_t=\{y\in{\mathbb M}_4: y^0=t\}$. The crucial issue is that the
Maxwell energy-momentum tensor density of entire system
\begin{equation}\label{T}
4\pi T^{\mu\nu} = f^{\mu\lambda}f^\nu{}_\lambda - 1/4\eta^{\mu\nu}
f^{\kappa\lambda}f_{\kappa\lambda}
\end{equation}
is the sum of individual "one-particle" densities and an "interference"
term:
\begin{equation}\label{T2}
T^{\mu\nu} = T_{(1)}^{\mu\nu}+T_{(2)}^{\mu\nu}+T_{\mbox{\scriptsize
int}}^{\mu\nu} .
\end{equation}
An intrigue feature is that the radiative contribution from the
combination of the retarded Li\'enard-Wiechert fields
\begin{equation}\label{T12}
4\pi T_{\mbox{\scriptsize int}}^{\mu\nu} =
f_{(1)}^{\mu\lambda}f_{(2)}^\nu{}_\lambda +
f_{(2)}^{\mu\lambda}f_{(1)}^\nu{}_\lambda - 1/4\eta^{\mu\nu} \left(
f_{(1)}^{\kappa\lambda}f_{\kappa\lambda}^{(2)} +
f_{(2)}^{\kappa\lambda}f_{\kappa\lambda}^{(1)} \right)
\end{equation}
is then nothing but the sum of work done by Lorentz forces of point-like
charges acting on one another. Therefore, an interference of outgoing
electromagnetic waves in an {\it observation hyperplane $\Sigma_t$} leads
to the interaction between the collided sources. (The
differentiation of energy-momentum conserved quantity gives the
relativistic generalization of Newton's second law \cite{RYar}.) This
observation gives us an alternative interpretation for the label "int":
it stands for "interaction" as well as "interference".

In this paper we study a closed system of two {\it arbitrarily moving}
point-like charges which are asymptotically free in the remote past. The
expressions for work done by (retarded) Lorentz forces will be obtained
via the rigorous integration of interference parts (\ref{T12}) of energy
and momentum densities (\ref{T2}) over three-dimensional hyperplane
$\Sigma_t$.

\section{Preliminaries}
\setcounter{equation}{0}

We choose metric tensor $\eta_{\mu\nu}={\rm diag}(-1,1,1,1)$ for
Minkowski space ${\mathbb M}_4$. We use Heaviside-Lorentz system of
units with the velocity of light $c=1$. Summation over repeated indices
is understood throughout the paper; Greek indices run from $0$ to $3$,
and Latin indices from $1$ to $3$. The particles' coordinates,
velocities etc are labelled $a$ or $b$.

We consider an arbitrarily moving particles which are asymptotically free
in the remote past. Average velocities are not large enough to initiate
particle creation and annihilation.

We suppose that the components of momentum four-vector carried by
electromagnetic field of particles are \cite{Rohr}
\begin{equation} \label{pem}
p_{em}^\nu (t)= P\int_{\Sigma_t}
d\sigma_\mu T^{\mu\nu} \,,
\end{equation}
where $d\sigma_\mu$ is the vectorial surface element on a {\it observation
hyperplane} $\Sigma_t=\{y\in{\mathbb M}_4: y^0=t\}$. Particles' world lines
\begin{eqnarray} \label{trj}
\zeta_a &:& {\mathbb R}
 \to {\mathbb M}_{\,4} \nonumber\\
&&t\mapsto (t,z_a^i(t))
\end{eqnarray}
are meant as local sections of trivial bundle $({\mathbb
M}_{\,4},i,\mathbb R)$  where the projection
\begin{eqnarray} \label{i}
i&:&{\mathbb M}_{\,4}\to \mathbb R \nonumber\\
&&(y^0,y^i)\mapsto y^0
\end{eqnarray}
defines the instant form of dynamics \cite{GKT}.

By $T^{\mu\nu}$ we denote the components of the Maxwell energy-momentum
density (\ref{T}) where field strengths $f^{\mu\nu}$ are the sum of the
retarded Li\'enard-Wiechert solutions $f_{(1)}^{\mu\nu}$ and
$f_{(2)}^{\mu\nu}$
associated with the first and second particles, respectively. So, the total
electromagnetic field stress-energy tensor (\ref{T}) becomes the sum
(\ref{T2}) where the $T_{(a)}^{\mu\nu}$ term is given by the expression
(\ref{T}) where "total" field strengths $f^{\mu\nu}$ are replaced by
"individual" ones $f_{(a)}^{\mu\nu}$. The interference term (\ref{T12})
describes the combination of the {\em outgoing} electromagnetic waves.

The components $T^{\mu\nu}$ have singularities on particles' trajectories.
In equations (\ref{pem}) capital letter $P$ denotes the principal value of
the singular integral, defined by removing from $\Sigma_t$ an
$\varepsilon_a$-sphere around the $a$-th particle and then
passing to the limit $\varepsilon_a\to 0$.

\section{"Interference" coordinate system}
\setcounter{equation}{0}
The main goal of the present paper is to compute the interference parts of
Poincar\'e group conserved quantities carried by radiation. To perform
the volume integration an appropriate coordinate system for flat
space-time is necessary.

\begin{figure}[t]
\begin{center}
\epsfclipon
\epsfig{file=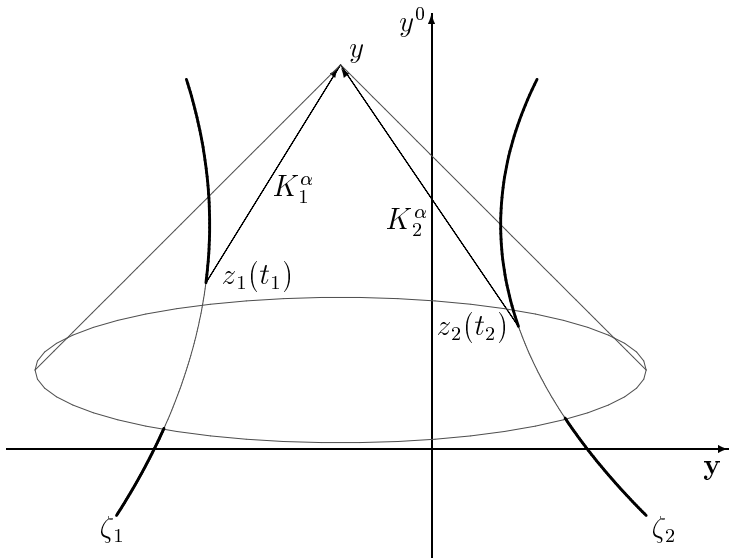,width=10cm}
\end{center}
\caption{\label{lc_map}
\small The past light cone with vertex at point $y\in\Sigma_t$ is
punctured by the world lines of the 1-st particle and the 2-nd particle at
points $z_1(t_1)$ and $z_2(t_2)$, respectively. The vector $K_a^\alpha$ is a
null vector pointing from $z_a(t_a)=(t_a,z_a^i(t_a))$ to $y$.
}
\end{figure}

\subsection{Local expressions}

The interference terms of energy-momentum and angular momentum at point
$y\in{\mathbb M}_4$ depend on the state of the charges' motion at the
instants $t_1$ and $t_2$ at which their world lines intersect the past
light cone (see Fig.\ref{lc_map}). Coordinates of an observation point $y$
are given by
\begin{equation} \label{y_a}
y^\alpha=z_a^\alpha(t_a)+K_a^\alpha
\end{equation}
where $K_a^\alpha$ is the null vector pointing from $z_a(t_a)\in\zeta_a$ to
$y$. Our next task is to find out local expressions for the "light-cone
mapping" \cite{Pois} pictured in Fig.\ref{lc_map}. We generalise coordinate
system presented in \cite{WFY} where a frontal collision is considered.

\begin{figure}[t]
\begin{center}
\epsfclipon
\epsfig{file=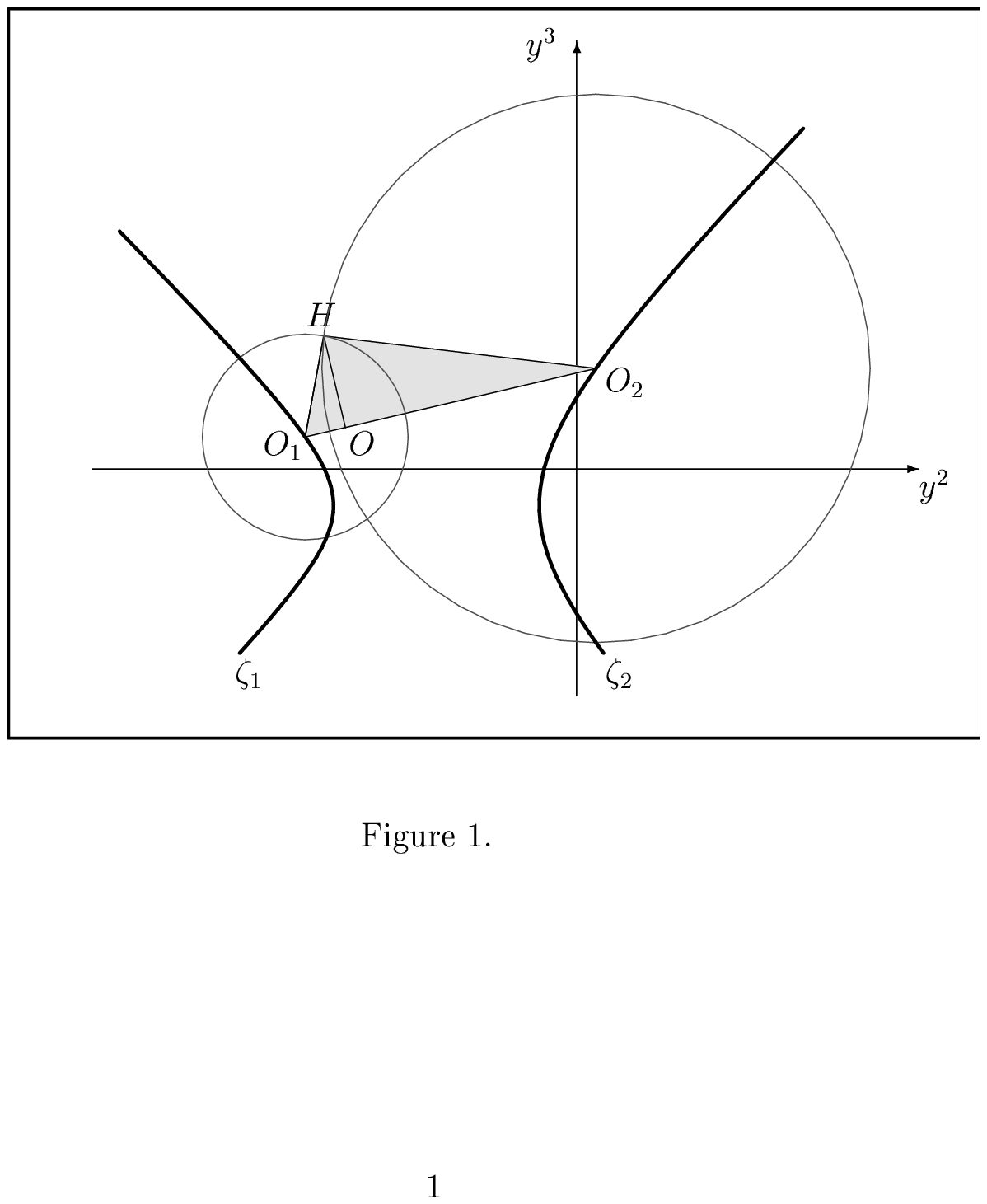,width=10cm}
\end{center}
\caption{\label{intf}
\small The sphere $S_1(O_1,t-t_1)$ is the intersection of the future light
cone with vertex at point $z_1(t_1)\in\zeta_1$ and hyperplane $\Sigma_t$.
The sphere $S_2(O_2,t-t_2)$ is the intersection of $\Sigma_t$ and the
forward light cone of $z_2(t_2)\in\zeta_2$. Intersection $S_1\cap S_2$ is
the circle $C(O,h)$ with radius $|OH|:=h$. It contains an observation point
$y\in\Sigma_t$ (see Fig.\ref{lc_map}).
}
\end{figure}

The set of curvilinear coordinates contains the "laboratory" time $t$
as well as both the "retarded" times $t_1$ and $t_2$. The "laboratory"
is a single common parameter defined along all the world lines of the
system. To find out local expressions for the components of null-vectors
$K_1$ and $K_2$ we consider an interference of outgoing electromagnetic
waves in hyperplane $\Sigma_t$ (see Fig.\ref{intf}). By this we mean the
intersection of spherical fronts $S_1(O_1,t-t_1)$ and $S_2(O_2,t-t_2)$
pictured in Fig.\ref{intf}.  It is the circle $C(O,h)$ centred at point
\begin{equation}\label{Z}
{\bf Z}=\frac12\Bigl[{\bf z}_1(t_1)+{\bf z}_2(t_2)\Bigr] +
\frac{(t_1-t_2)(2t-t_1-t_2)}{2q^2}\Bigl[{\bf z}_1(t_1)-{\bf
z}_2(t_2)\Bigr] .
\end{equation}
Since $|O_1O|=|{\bf Z}-{\bf z}_1|$ and $|OO_2|=|{\bf Z}-{\bf z}_2|$, the
square of the radius $h$ of the circle can be expressed in the following
alternative ways: \begin{eqnarray} \label{h}
h^2&=(t-t_1)^2-|{\bf Z}-{\bf z}_1|^2 \nonumber\\
&=(t-t_2)^2-|{\bf Z}-{\bf z}_2|^2 .
\end{eqnarray}
The characteristics of the circle are obtained from analysis of the triangle
$O_1O_2H$ with sides $|O_1H|=t-t_1$, $|O_2H|=t-t_2$, and
$|O_1O_2|=|{\bf z}_1(t_1)-{\bf z}_2(t_2)|:=q$.

\begin{figure}[h]
\begin{center}
\epsfclipon
\epsfig{file=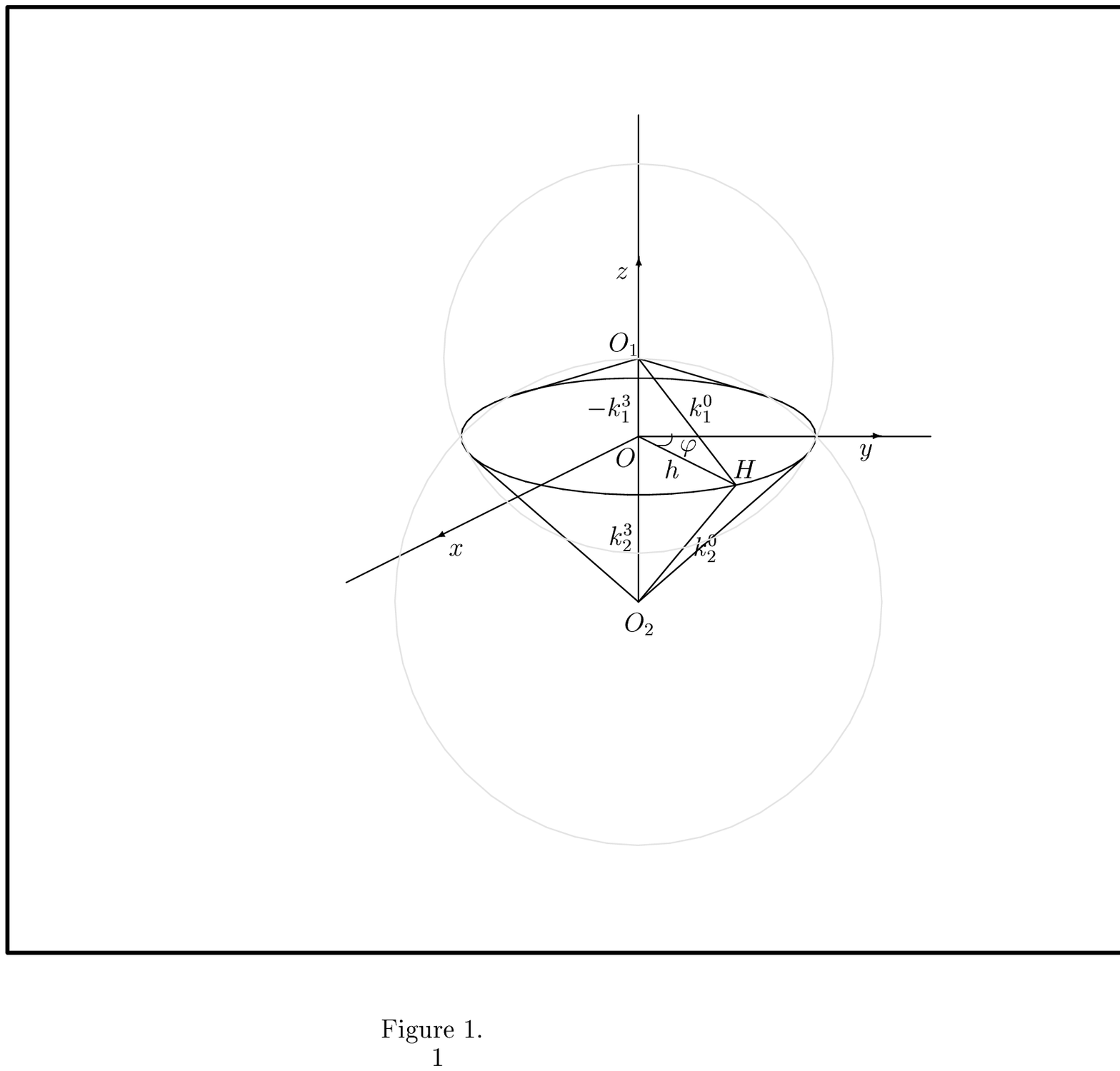,width=7cm}
\end{center}
\caption{\label{k_k}
\small In "momentarily rotating" Lorentz frame $z-$axis is directed along
three-vector ${\bf q}$. Circle $C(O,h)=S_1\cap S_2$ lies in $Oxy$ plane; it
is centred at the coordinate origin (cf. Fig.\ref{intf}). Polar angle
$\varphi$ distinguishes
an observation point $H\in C(O,h)$. Space parts ${\bf k}_1$ and ${\bf k}_2$
of null vectors $k_1$ and $k_2$ are equal to $h\sin\varphi{\bf i}+
h\cos\varphi{\bf j}+k_1^3{\bf k}$ and $h\sin\varphi{\bf i}+
h\cos\varphi{\bf j}+k_2^3{\bf k}$, respectively.
}
\end{figure}

To define the coordinates of the points of the circle we translate the
origin at the centre (\ref{Z}) of the circle $C(O,h)$ and then rotate space
axes till new $z$-axis be directed along three-vector ${\bf q}:={\bf
z}_1-{\bf z}_2$ (see Fig.\ref{k_k}). Orthogonal matrix
\begin{equation}\label{om}
\omega=\left(
\begin{array}{ccc}
\cos\varphi_q&-\sin\varphi_q&0\\
\sin\varphi_q&\cos\varphi_q&0\\
0 & 0 & 1
\end{array}
\right)
\left(
\begin{array}{ccc}
\cos\vartheta_q&0&\sin\vartheta_q\\
0 & 1 & 0\\
-\sin\vartheta_q&0&\cos\vartheta_q
\end{array}
\right)
\end{equation}
determines the rotation. Finally we obtain coordinate transformation
locally written as
\begin{eqnarray} \label{Zh}
y^0&=&t\nonumber\\
y^i&=&Z^i(t,t_1,t_2)+h(t,t_1,t_2)\omega^i{}_j(t_1,t_2)n^j
\end{eqnarray}
where $n^j=(\sin\varphi,\cos\varphi,0)$. Polar angle $\varphi$
distinguishes the points of circle $C(O,h)$.

To present the local expressions for the {\em a coordinate system centred
on an accelerated world line} of the a-th particle, we rewrite
eqs.(\ref{Zh}) in a manifestly covariant fashion:
\begin{equation} \label{Zha}
y^\alpha=z_a^\alpha(t_a)+\Omega^\alpha{}_{\alpha'}(t_1,t_2)k_a^{\alpha'} .
\end{equation}
Four components
\begin{equation} \label{kk}
k_a^0=t-t_a,\quad k_a^1=h\sin\varphi, \quad k_a^2=h\cos\varphi,
\quad k_a^3=(-1)^a|{\bf Z} - {\bf z}_a|
\end{equation}
satisfy the relations ($\ref{h}$) and, therefore, constitute null-vector
$k_a$. Having rotated it by orthogonal matrix $\Omega$ with components
$\Omega_{0\mu}=\Omega_{\mu 0}=\delta_{\mu 0}, \Omega_{ij}=\omega_{ij}$
we obtain the vector $K_a$ pointing from $z_a(t_a)\in\zeta_a$ to
$y\in\Sigma_t$ (see Fig.\ref{lc_map}). The orthogonal matrix
$\omega$ is given by eq.(\ref{om}); it rotates space axes of the
laboratory Lorentz frame (see Fig.\ref{k_k}).

Third component of $k_a$ is determined by
\begin{equation}\label{Zza}
|{\bf Z}-{\bf z_a}|=\frac{\displaystyle q}{\displaystyle 2} +
(-1)^a\frac{\displaystyle (k_2^0)^2-(k_1^0)^2}{\displaystyle
2q}.
\end{equation}
The characteristics $|{\bf Z}-{\bf z_1}|$ and $|{\bf Z}-{\bf z_2}|$ are
obtained from the analysis of the triangle $O_1O_2H$ with sides
$|O_1H|=t-t_1$, $|O_2H|=t-t_2$ and $|O_1O_2|=|{\bf z_1}(t_1)-{\bf
z_2}(t_2)|:=q$; they are pictured in Fig.(\ref{k_k}).

\subsection{Global mapping}

\begin{figure}[t]
\begin{center}
\epsfclipon
\epsfig{file=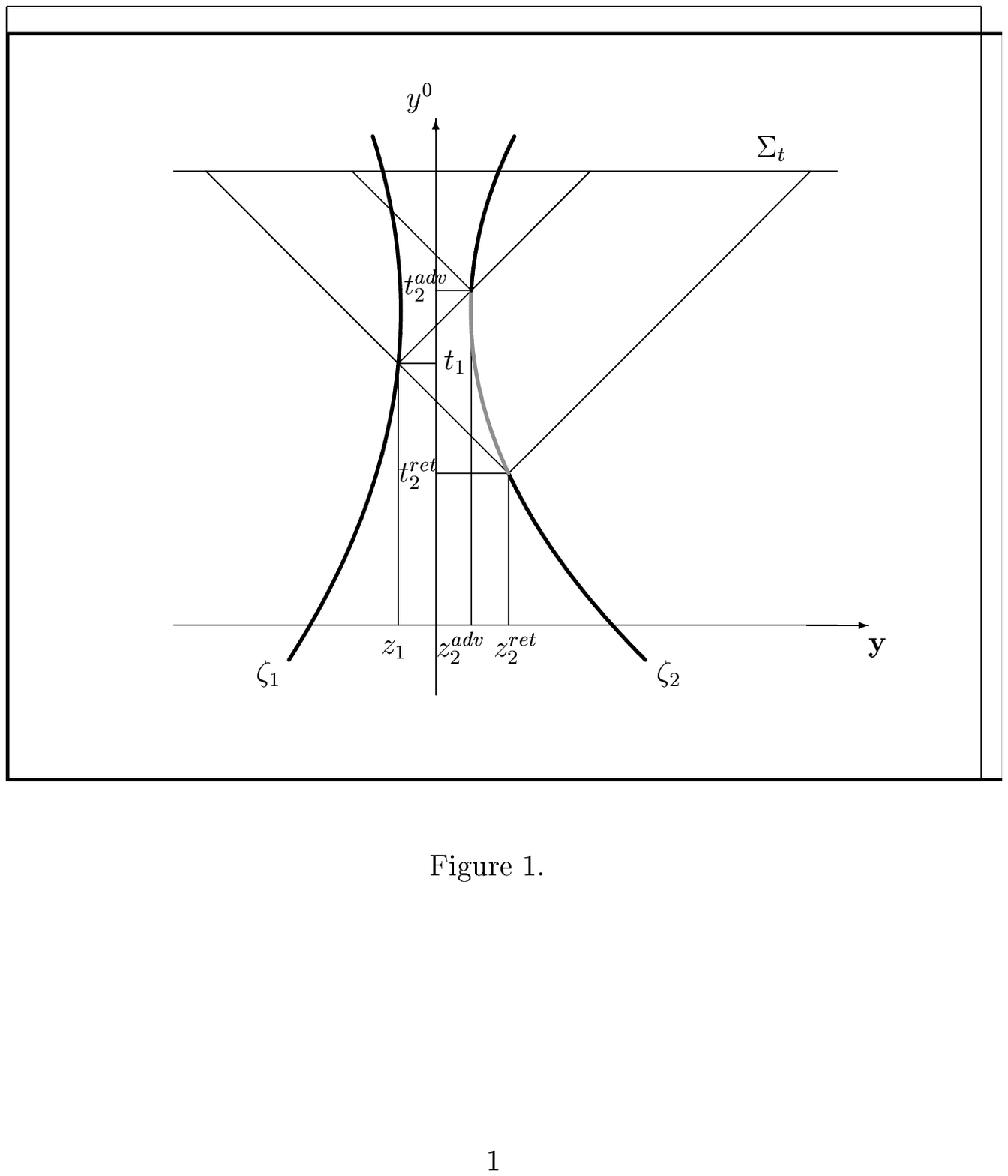,width=10cm}
\end{center}
\caption{\label{ret_adv}
\small For a given $t_1$ the retarded time $t_2$ increases from
$t_2^{ret}(t_1)$ to $t_2^{adv}(t_1)$.
Minimal value $t_2^{ret}(t_1)$ labels the vertex of
forward light cone which is punctured by the world line of the first
charge at a given point $(t_1,z_1^i(t_1))$. The world line of the second
charge punctures the future light cone of this point at point
$(t_2^{adv}(t_1),z_2^i(t_2^{adv}))$.
}
\end{figure}

\begin{figure}[h]
\begin{center}
\epsfclipon
\epsfig{file=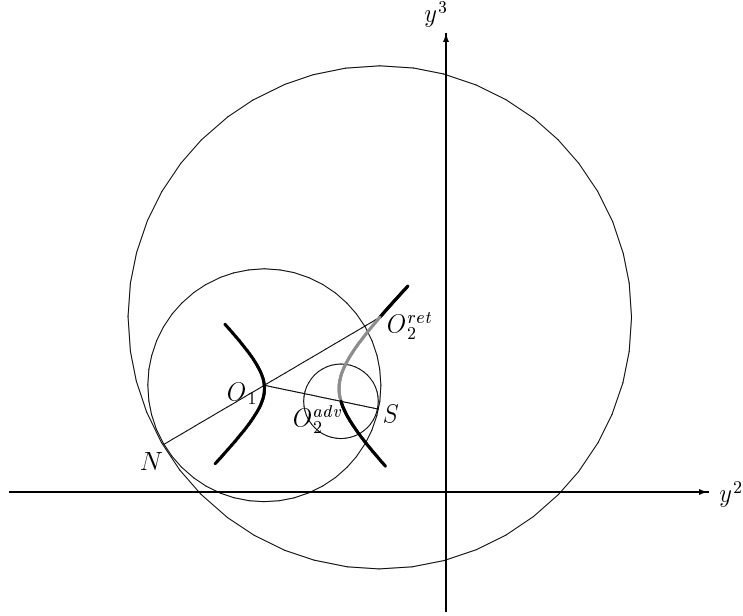,width=10cm}
\end{center}
\caption{\label{ra_cr}
\small The sphere $S_2(O_2^{ret},t-t_2^{ret})$ is the intersection of the
future light cone at $(t_2^{ret},z_2^i(t_2^{ret}))$ and
$\Sigma_t$. It touches a given sphere $S_1(O_1,t-t_1)$ at point $N$. The
sphere
$S_2(O_2^{adv},t-t_2^{adv})$ touches $S_1(z_1,t-t_1)$ at point $S$.
If retarded time $t_2$ increases from $t_2^{ret}(t_1)$ to
$t_2^{adv}(t_1)$ the sphere $S_1$ is covered by circles $C(O,h)=S_1\cap
S_2$. (A circle $S_1\cap S_2$ is pictured in Figs.\ref{intf},\ref{k_k}.)
}
\end{figure}

To cover the sphere $S_1(z_1(t_1),t-t_1)$ where $t_1$ is fixed we change
the parameter $t_2$. The starting point is the solution $t_2^{ret}(t_1)$
of algebraic equation
\begin{equation}\label{ret2}
t_1-t_2^{ret}=q(t_1,t_2^{ret})
\end{equation}
which describes the future light cone with vertex at
$(t_2^{ret},z_2^i(t_2^{ret}))$ (see Fig.\ref{ret_adv}).
The sphere $S_2(z_2^{ret},t-t_2^{ret})$ touches a given sphere
$S_1(z_1(t_1),t-t_1)$ at point N (see Fig.\ref{ra_cr}).
If parameter $t_2$ increases to $t_2^{adv}(t_1)$
being the solution of algebraic equation
\begin{equation}\label{adv2}
t_2^{adv}-t_1=q(t_1,t_2^{adv})
\end{equation}
the intersection $S_1\cap S_2^{adv}$ contains the only point S.
Equation (\ref{adv2}) looks as the equation of {\it backward} light cone of
$(t_2^{adv},z_2^i(t_2^{adv}))$, but it defines the {\it future} light cone
with vertex at $(t_1,z_1^i(t_1))$ (see Fig.\ref{ret_adv}). The sphere $S_1$
becomes the disjoint union of circles $C(O,h)=S_1\cap S_2$ if the parameter
$t_2$ changes from $t_2^{ret}(t_1)$ to $t_2^{adv}(t_1)$.

\begin{figure}[h]
\begin{center}
\epsfclipon
\epsfig{file=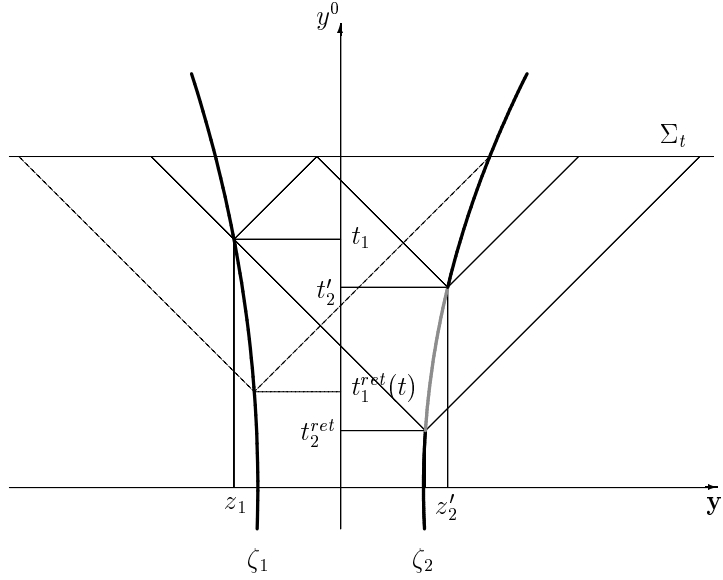,width=10cm}
\end{center}
\caption{\label{n-caus}
\small The forward light cone of $(t_1^{ret}(t),z_1^i(t_1^{ret}))$ touches
the second world line at the instant of observation. Future light cones of
upper vertices do not intersect it at all. For a given $t_1\in
[t_1^{ret}(t),t]$ the parameter $t_2$ increases from
$t_2^{ret}(t_1)$ to $t_2'(t,t_1)$. The maximal value $t_2'(t,t_1)$ labels
the vertex of future light cone which touches the forward light cone of
$(t_1,z_1^i(t_1))$. The minimal value of $t_2$ is the solution
$t_2^{ret}(t_1)$ of equation (\ref{ret2}).
}
\end{figure}

\begin{figure}[h]
\begin{center}
\epsfclipon
\epsfig{file=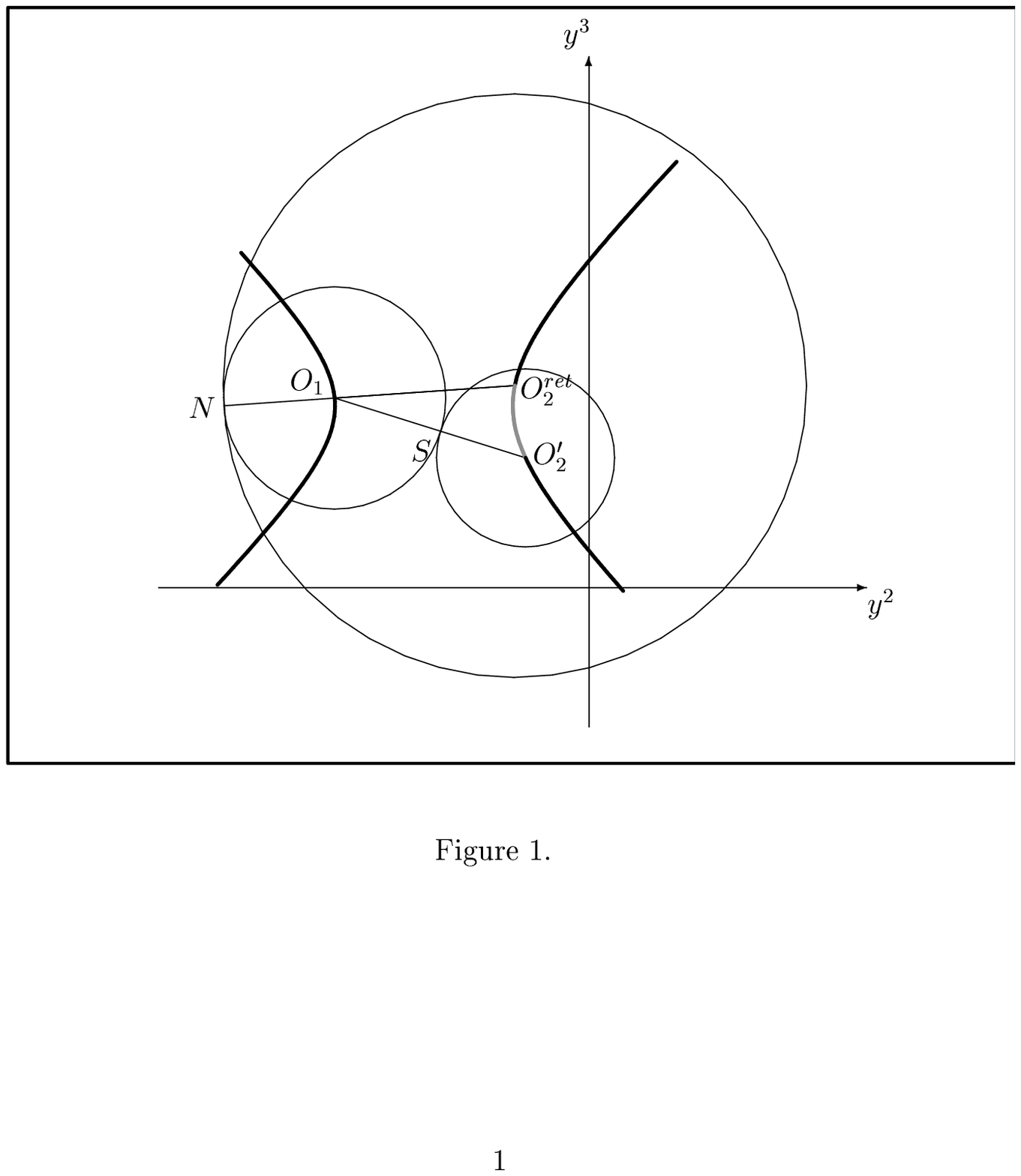,width=10cm}
\end{center}
\caption{\label{nc-cr}
\small For a given $t_1\in [t_1^{ret}(t),t]$ the sphere $S_1(O_1,t-t_1)$ is
a disjoint union of circles $C(O,h)=S_1\cap S_2$. Their radius $h$ and
centre coordinate $Z$ are determined by $t_2$. The parameter $t_2$
increases from $t_2^{ret}(t_1)$ (circle $S_2(O_2^{ret},t-t_2^{ret})$) to
$t_2'(t,t_1)$ (circle $S_2(O_2',t-t_2')$); $\varphi\in [0,2\pi ]$.
}
\end{figure}

Going along the world line of the first charge we arrive unavoidably at
the point $t_1^{ret}(t)$ being the solution of the algebraic equation
\begin{equation}\label{t1ret}
t-t_1^{ret}=q(t_1^{ret},t).
\end{equation}
The forward light cone of this point touches the world line of second
charge at point $(t,z_2^i(t))$ (see Fig.\ref{n-caus}).
Light cones of upper vertices do not intersect the second world line at
all. Spheres $S_1(z_1(t_1),t-t_1)$ determined by $t_1\in
[t_1^{ret}(t),t]$ constitute the region of hyperplane $\Sigma_t$ which
requires another parametrization. For a given instant $t_1$ from this
interval the point $S$ (see Fig.\ref{nc-cr})
is associated with the solution $t_2'(t_1)$ of the following equation:
\begin{equation}\label{t2prime}
2t-t_1-t_2'=q(t_1,t_2')\,.
\end{equation}
The point $N$ in this figure is still connected with the solution
$t_2^{ret}(t_1)$ of equation (\ref{ret2}).

So, we construct the global coordinate system centred on the world line
of the first particle. It bases on the trivial fibre bundle (\ref{i}).
A fibre $\Sigma_t$ is a disjoint union of retarded spheres $S_1$ centred
on the world line of the first particle. A sphere is parametrized by the
retarded time of the second particle and the polar angle. Locally the
coordinate transformation is given by equations (\ref{Zh}).

In an analogous way we construct the coordinate system centred on the
world line of the second particle. If $t_2\in ]-\infty,t_2^{ret}(t)]$ then
$t_1\in [t_1^{ret}(t_2), t_1^{adv}(t_2)]$; if $t_2\in
[t_2^{ret}(t),t]$ then $t_1\in [t_1^{ret}(t_2),
t_1'(t,t_2)]$, $\varphi \in [0,2\pi [$. The ends of intervals are
defined by the following algebraic equations:
\begin{eqnarray}
t_2-t_1^{ret}&=q(t_1^{ret},t_2) \\ \label{ret1}
t_1^{adv}-t_2&=q(t_1^{adv},t_2)\\ \label{adv1}
t-t_2^{ret}&=q(t,t_2^{ret}) \\ \label{t2ret}
2t-t_1'-t_2&=q(t_1',t_2)\,. \label{t1prime}
\end{eqnarray}
It is worth noting that the functions $t_1^{ret}(t_2)$
and $t_2^{adv}(t_1)$  are inverted to each other as
well as the pair of functions $t_1^{adv}(t_2)$
and $t_2^{ret}(t_1)$ (see Fig.\ref{A-B}). For a fixed observation time
$t$ the functions $t_1'(t,t_2)$  and $t_2'(t,t_1)$ are
inverses too.

\section{Electromagnetic fields in terms of "interference" coordinates}
\setcounter{equation}{0}

Electromagnetic field generated by $a-$th particle is given by
\cite[eq.(5.2)]{Pois}
\begin{equation}\label{fP}
f_{\alpha\beta}^{(a)}=e_a\frac{\displaystyle
u_{a,\alpha}{\sf k}_{a,\beta}-u_{a,\beta}{\sf
k}_{a,\alpha}}{\displaystyle ({\sf r}_a)^2}\left[
1+{\sf r}_a({\sf k}_a\cdot a_a)\right] +
e_a\frac{\displaystyle
a_{a,\alpha}{\sf k}_{a,\beta}-a_{a,\beta}{\sf
k}_{a,\alpha}}{\displaystyle {\sf r}_a}.
\end{equation}
We use {\sf sans-serif} symbols for the retarded distance
\cite{Pois,Pois1}
\begin{equation}\label{R}
{\sf r}_a(y) = -\eta_{\alpha\beta}(y^\alpha -z^\alpha(t_a))u^\beta(t_a),
\end{equation}
and for the null vector $K_a$ rescaled by a factor ${\sf r}_a^{-1}$:
\begin{equation}\label{k_a}
{\sf k}_a^\alpha =\frac{\displaystyle 1}{\displaystyle {\sf r}_a}
\left[y^\alpha - z_a^\alpha(t_a)\right].
\end{equation}
To rewrite expression (\ref{fP}) in terms of "interference"
curvilinear coordinates consisting of the common evolution parameter $t$,
individual times $t_1$ and $t_2$, and angle variable $\varphi$, it is
advantageous to replace proper time $\tau_a$ by evolution parameter $t_a$.
The components of particles' 4-velocities $u_a$ and 4-accelerations $a_a$,
$a=1,2$, become \cite{Rohr}
\begin{eqnarray}\label{u_a}
u_a^\mu&=&\gamma_av_a^\mu(t_a)\\
a_a^\mu&=&\gamma_a^4(v_a\cdot{\dot v}_a)v_a^\mu+\gamma_a^2{\dot v}_a^\mu
\end{eqnarray}\label{a_a}
where 4-vectors $v_a^\mu=(1,v_a^i(t_a))$, ${\dot
v}_a^\mu=(0,{\dot v}_a^i(t_a))$ and factor $\gamma_a:=[1-{\bf
v}^2]^{-1/2}$.
After some algebra, using the relation ${\sf k}_a^\mu=K_a^\mu/{\sf
r}_a$, we obtain
\begin{eqnarray}\label{f}
f_{\alpha\beta}^{(a)}&=&e_a\frac{v_\alpha(t_a)K_{a,\beta}-
v_\beta(t_a)K_{a,\alpha}}{r_a^3}c_a\nonumber\\
&+&e_a\frac{{\dot v}_\alpha(t_a)K_{a,\beta}-
{\dot v}_\beta(t_a)K_{a,\alpha}}{r_a^2}
\end{eqnarray}
where
\begin{equation}\label{cr}
c_a=\gamma_a^{-2}+({\bf K}_a{\bf\dot v}_a),\qquad
r_a=K_a^0 - ({\bf K}_a{\bf v}_a).
\end{equation}

Having used differential chart (\ref{diff}), one can derive the
electromagnetic field (\ref{f}) from Li\'enard-Wiechert potential
\begin{equation}
A_\alpha^{(a)} = e_a\frac{\displaystyle u_\alpha(t_a)}{\displaystyle
{\sf r}_a(y)}
\end{equation}
via the relations
$f_{\alpha\beta}^{(a)}=A_{\beta,\alpha}^{(a)}-A_{\alpha,\beta}^{(a)}$.

\section{Interference part of electromagnetic field\protect\\
four-momentum}
\setcounter{equation}{0}
Now, we calculate the interference part of the energy and momentum
carried by "two-particle" electromagnetic field:
\begin{equation} \label{pint}
p_{\mbox{\scriptsize int}}^\mu (t)=\int_{\Sigma_t}
d\sigma_0 T_{\mbox{\scriptsize int}}^{0\mu} \,.
\end{equation}
An integration hypersurface $\Sigma_t=\{y\in {\mathbb M}_4:
y^0=t\}$ is a surface of constant $t$. The surface element is given by
$d\sigma_0=\sqrt{-g}d t_1d t_2d\varphi$ where $\sqrt{-g}$
is the determinant of metric tensor of Minkowski space viewed in
curvilinear coordinates (\ref{Zh}). Differentiation of coordinate
transformation (\ref{Zh}) yields differential chart
\begin{eqnarray}\label{diff}
\left(
\begin{array}{c}
\partial/\partial y^0\\[10pt]
\partial/\partial y^1\\[10pt]
\partial/\partial y^2\\[10pt]
\partial/\partial y^3
\end{array}
\right)&=&
\left(
\begin{array}{cccc}
1 & 0 & 0 & 0\\[10pt]
0 &\omega_{11} & \omega_{12} & \omega_{13} \\[10pt]
0 &\omega_{21} & \omega_{22} & \omega_{23} \\[10pt]
0 &\omega_{31} & \omega_{32} & \omega_{33}
\end{array}
\right)\\
&&\left(
\begin{array}{cccc}
1 &  \frac{\displaystyle t-t_1}{\displaystyle r_1}& \frac{\displaystyle
t-t_2}{\displaystyle r_2} & 0\\[5pt]
0 & -\frac{\displaystyle h\sin\varphi}{\displaystyle r_1} &
-\frac{\displaystyle h\sin\varphi}{\displaystyle r_2} &
\frac{\displaystyle \cos\varphi}{\displaystyle h} \\[5pt]
0 & -\frac{\displaystyle h\cos\varphi}{\displaystyle r_1} &
-\frac{\displaystyle h\cos\varphi}{\displaystyle r_2} &
-\frac{\displaystyle \sin\varphi}{\displaystyle h} \\[5pt]
0 & \frac{\displaystyle |{\bf Z}-{\bf z_1}|}{\displaystyle r_1}  &
-\frac{\displaystyle |{\bf Z}-{\bf z_2}|}{\displaystyle r_2}  &  0
\end{array}
\right)
\left(
\begin{array}{c}
\partial/\partial t\\[10pt]
\partial/\partial t_1\\[10pt]
\partial/\partial t_2\\[10pt]
\partial/\partial \varphi
\end{array}
\right) \nonumber
\end{eqnarray}
Its Jacobian gives the determinant of metric tensor mentioned above
\begin{equation} \label{se}
\sqrt{-g} = \frac{\displaystyle r_1r_2}{\displaystyle q}.
\end{equation}

The volume integration (\ref{pint}) can be performed via the coordinate
system centred on a world line either of the first particle
\begin{equation} \label{int1}
\left[\int\limits_{-\infty}^{t_1^{ret}(t)}d  t_1
\int\limits_{t_2^{ret}(t_1)}^{t_2^{adv}(t_1)}d  t_2
+ \int\limits_{t_1^{ret}(t)}^td  t_1
\int\limits_{t_2^{ret}(t_1)}^{t_2'(t,t_1)}d  t_2
\right]\int_0^{2\pi}d \varphi\frac{r_1r_2}{q}
\end{equation}
or of the second particle
\begin{equation} \label{int2}
\left[\int\limits_{-\infty}^{t_2^{ret}(t)}d  t_2
\int\limits_{t_1^{ret}(t_2)}^{t_1^{adv}(t_2)}d  t_1
+ \int\limits_{t_2^{ret}(t)}^td  t_2
\int\limits_{t_1^{ret}(t_2)}^{t_1'(t,t_2)}d  t_1
\right]\int_0^{2\pi}d \varphi\frac{r_1r_2}{q}\,.
\end{equation}
The end points of these integrals arise from the interference pictured in
Figs.\ref{intf}-\ref{nc-cr}.

\subsection{Interference part of zeroth component}

In this subsection we trace a series of stages in calculation of the
volume integral
\begin{equation} \label{p0int}
p_{\mbox{\scriptsize int}}^0=\int_{\Sigma_t}
d\sigma_0 T_{\mbox{\scriptsize int}}^{00} \,.
\end{equation}
In Appendix A we perform the computation in detail.

It is straightforward to substitute the components (\ref{f}) into
equation (\ref{T12})
to calculate the interference part of electromagnetic
field stress-energy tensor. We obtain the following energy density:
\begin{equation}\label{T00}
4\pi T_{\mbox{\scriptsize int}}^{00}=\frac{\displaystyle e_1e_2}{\displaystyle
r_1r_2}\left(
\frac{\displaystyle \partial^2\Gamma_0}{\displaystyle \partial t_1\partial
t_2}\frac{\displaystyle 1}{\displaystyle r_1r_2}+
\frac{\displaystyle \partial\Gamma_0}{\displaystyle \partial
t_1}\frac{\displaystyle c_2}{\displaystyle r_1(r_2)^2} +
\frac{\displaystyle \partial\Gamma_0}{\displaystyle \partial
t_2}\frac{\displaystyle c_1}{\displaystyle (r_1)^2r_2} +
\Gamma_0\frac{\displaystyle c_1c_2}{\displaystyle (r_1)^2(r_2)^2}
\right)
\end{equation}
where function
\begin{equation}
\Gamma_0 = \kappa\frac{\displaystyle \partial^2\kappa}{\displaystyle
\partial t_1\partial t_2}-
\frac{\displaystyle \partial\kappa}{\displaystyle \partial t_1}
\frac{\displaystyle \partial\kappa}{\displaystyle \partial t_2},
\qquad
\kappa = \frac12(k_1^0+k_2^0)^2-\frac12q^2
\end{equation}
does not depend on angle variable at all.

Taking into account the specific structure of the expression (\ref{T00})
which contains the partial derivatives we rewrite the integrand
$\sqrt{-g}T_{\bf int}^{00}$ as follows:
\begin{eqnarray}\label{JT00}
\frac{4\pi}{e_1e_2}\frac{r_1r_2}{q}T_{\rm
int}^{00}&=&\frac{\partial^2}{\partial t_1\partial
t_2}\left(\frac{\Gamma_0}{qr_1r_2}\right) +
\frac{\partial}{\partial
t_1}\left\{\Gamma_0\left[\frac{c_2}{qr_1(r_2)^2}-\frac{\partial}{\partial
t_2}\left(\frac{1}{qr_1r_2}\right)\right]\right\}
\nonumber\\
&+&\frac{\partial}{\partial
t_2}\left\{\Gamma_0\left[\frac{c_1}{q(r_1)^2r_2}-\frac{\partial}{\partial
t_1}\left(\frac{1}{qr_1r_2}\right)\right]\right\}
\nonumber\\
&+&\Gamma_0\left[\frac{c_1c_2}{q(r_1)^2(r_2)^2} -
\frac{\partial}{\partial
t_1}\left(\frac{c_2}{qr_1(r_2)^2}\right) -
\frac{\partial}{\partial
t_2}\left(\frac{c_1}{q(r_1)^2r_2}\right) \right.\nonumber\\
&+&\left.
\frac{\partial^2}{\partial
t_1\partial t_2}\left(\frac{1}{qr_1r_2}\right)
\right].
\end{eqnarray}
First of all we should perform the integration over $\varphi$ (see
integration rules (\ref{int1}) and (\ref{int2})). The crucial issue is
that the integral of the bracketed expression (that which is proportional
to $\Gamma_0$) over $\varphi$ vanishes (see Appendix A).
Hence the integral of (\ref{JT00}) over the angle variable has the
remarkable properties of being the sum of partial derivatives:
\begin{eqnarray}\label{fiT00}
\int\limits_0^{2\pi} d\varphi \frac{r_1r_2}{q}T_{\rm
int}^{00}&=&\frac{e_1e_2}{2}\left\{\frac{\partial^2
(\Gamma_0{\cal D}_0)}{\partial t_1\partial t_2} +
\frac{\partial}{\partial t_1}\left[\Gamma_0\left({\cal
B}_0-\frac{\partial {\cal D}_0}{\partial t_2}\right)\right]\right.
\nonumber\\
&+& \left.\frac{\partial}{\partial
t_2}\left[\Gamma_0\left({\cal C}_0-\frac{\partial {\cal D}_0}{\partial
t_1}\right)\right]\right\}.
\end{eqnarray}
Here
\begin{equation}\label{BCD0}
{\cal D}_0=\frac{1}{2\pi}\int\limits_0^{2\pi} d\varphi
\frac{1}{qr_1r_2},\quad
{\cal B}_0=\frac{1}{2\pi}\int\limits_0^{2\pi} d\varphi \frac{c_2}{qr_1(r_2)^2},\quad
{\cal C}_0=\frac{1}{2\pi}\int\limits_0^{2\pi} d\varphi \frac{c_1}{q(r_1)^2r_2}
\end{equation}
where $r_a$ and $c_a$ are given by eqs.(\ref{cr}).

It is natural to integrate the expression being the time derivative with
respect to $t_1$ according to the rule (\ref{int2}). The result is
\begin{eqnarray} \label{G2}
&&\left[\int\limits_{-\infty}^{t_1^{ret}(t)}d  t_1
\int\limits_{t_2^{ret}(t_1)}^{t_2^{adv}(t_1)}d  t_2
+ \int\limits_{t_1^{ret}(t)}^td  t_1
\int\limits_{t_2^{ret}(t_1)}^{t_2'(t,t_1)}d  t_2
\right]\frac{\partial G_2(t_1,t_2)}{\partial t_2}\\
&=&\int\limits_{-\infty}^{t_1^{ret}(t)}d  t_1G_2[t_1,t_2^{adv}(t_1)] -
\int\limits_{-\infty}^{t}d t_1G_2[t_1,t_2^{ret}(t_1)] +
\int\limits_{t_1^{ret}(t)}^td t_1G_2[t_1,t_2'(t,t_1)].\nonumber
\end{eqnarray}
Having applied the rule (\ref{int1}) to the expression of type $\partial
G_1/\partial t_1$, we obtain
\begin{eqnarray} \label{G1}
&&\left[\int\limits_{-\infty}^{t_2^{ret}(t)}d  t_2
\int\limits_{t_1^{ret}(t_2)}^{t_1^{adv}(t_2)}d  t_1
+ \int\limits_{t_2^{ret}(t)}^td  t_2
\int\limits_{t_1^{ret}(t_2)}^{t_1'(t,t_2)}d  t_1
\right]\frac{\partial G_1(t_1,t_2)}{\partial t_1}\\
&=&\int\limits_{-\infty}^{t_2^{ret}(t)}d t_2 G_1[t_1^{adv}(t_2),t_2]
-\int\limits_{-\infty}^{t}d t_2 G_1[t_1^{ret}(t_2),t_2]
+\int\limits_{t_2^{ret}(t)}^td t_2 G_1[t_1'(t,t_2),t_2]\nonumber
\end{eqnarray}

The double derivative involved in eq.(\ref{fiT00}) can be written in the
form either
\begin{equation} \label{dd1}
\frac{\partial}{\partial t_1}\left[\frac{\partial G_0}{\partial
t_2}\right]
\end{equation}
or
\begin{equation} \label{dd2}
\frac{\partial}{\partial t_2}\left[\frac{\partial G_0}{\partial
t_1}\right]\,.
\end{equation}
Now we choose (\ref{dd1}) and add this term to $\partial
G_1/\partial t_1$.

Therefore, the end points are valuable only in the integration procedure
either (\ref{int1}) or (\ref{int2}). The retarded instant,
$t_a^{ret}(t_b)$, and advanced one, $t_a^{ret}(t_b)$, ($a\ne b$) arise
naturally as the limits of integrals. They label the points $S$ and $N$
in which fronts of outgoing electromagnetic waves produced by $e_1$ and
$e_2$ touch each other (see Figs.\ref{ra_cr} and \ref{nc-cr}).
Triangle $O_1O_2H$ (see Fig.\ref{k_k}) reduces to the line at these
moments.

An essential feature of integration is that the functions $t_1^{adv}(t_2)$
and $t_2^{ret}(t_1)$ are inverted to each other (see Fig.\ref{A-B}). This
cicumstance allows us to change the variables in the "advanced" integral
involved in eq.(\ref{G1}). Further we couple it with the "retarded"
integral of eq.(\ref{G2}). We obtain
\begin{equation}\label{rt1}
\int\limits_{-\infty}^{t}d
t_1\left[\frac{1-V_1}{1-V_2}G_1-G_2\right]_{t_2=t_2^{ret}(t_1)}
\end{equation}
where
\begin{equation}
V_a:=({\bf n}_q{\bf v}_a).
\end{equation}
Scrupulous calculation results the terms of two quite different types:
(i) this depends on all previous evolution of the 1-st charge
\begin{equation} \label{10ye}
-\int\limits_{-\infty}^{t}d t_1\gamma_1^{-1}F^0_{21}[t_1,t_2^{ret}(t_1)]
\end{equation}
(ii) those determinated by the state of particles' motion at the
observation instant only:
\begin{eqnarray}\label{1ye}
&&e_1e_2\left[\frac{1+V_2}{2[t-t_2^{ret}(t_1)](1-V_2)}-
\frac{1}{q[t_1,t_2^{ret}(t_1)](1-V_2)}\right]_{t_1\to
-\infty}^{t_1=t}\nonumber\\
&=&-\frac{e_1e_2}{2[t-t_2^{ret}(t)]}
\end{eqnarray}
(see Appendix E, Table 1, left column, first line). The integral
(\ref{10ye}) over world line $\zeta_1$ is then nothing but the zeroth
component of the work done by "retarded" Lorentz force acting on the first
charge.

It is reasonable that, starting with the retarded Li\'enard-Wiechert
solutions, we obtain the retarded direct field due the 2-nd charge on the
1-st one. A surprising feature is that we can arrive at the expression for
the {\em advanced} direct field within the framework of retarded causality.
E.g., one can perform change of variables $(t_1^{ret}(t_2),t_2)\mapsto
(t_1,t_2^{adv}(t_1))$ in the {\em retarded} integral involved in
eq.(\ref{G1}) and then couple it with the {\em advanced} expression from
eq.(\ref{G2}):
\begin{equation}\label{ad1}
\int\limits_{-\infty}^{t_1^{ret}(t)}d
t_1\left[G_2-\frac{1+V_1}{1+V_2}G_1\right]^{t_2=t_2^{adv}(t_1)} .
\end{equation}
Having integrated (\ref{ad1}), we obtain the work done by {\em advanced}
Lorentz force due to the 2-nd charge plus functions of momentary positions
of particles:
\begin{equation}\label{10blue}
-\int\limits_{-\infty}^{t_1^{ret}(t)}d
t_1\gamma_1^{-1}F^0_{21}[t_1,t_2^{adv}(t_1)] +
\left[\frac{\displaystyle 1}{\displaystyle 2k_2^0}\frac{\displaystyle
1-V_2}{\displaystyle 1+V_2}
+\frac{\displaystyle 1}{\displaystyle q[1+V_2]}
\right]_{t_1\to -\infty}^{t_1\to t_1^{ret}(t)}
\end{equation}
(see Appendix E, Table 1, left column, second line). The matter is that
the integral of advanced force due to 2-nd charge over worldline
$\zeta_1$ is intimately connected with integral of the retarded force due
to 1-st charge over $\zeta_2$:
\begin{eqnarray}\label{1blue}
&&\int\limits_{-\infty}^{t}d
t_2\gamma_2^{-1}F^0_{12}[t_1^{ret}(t_2),t_2]
-\int\limits_{-\infty}^{t_1^{ret}(t)}d
t_1\gamma_1^{-1}F^0_{21}[t_1,t_2^{adv}(t_1)]\\
&=&-e_1e_2\left[\frac{-1+({\bf v}_1\cdot{\bf v}_2)}{q[1+V_1][1+V_2]}
+\frac{1}{q[1+V_1]}+\frac{1}{q[1+V_2]}
\right]_{t_2\to -\infty}^{t_2=t}\nonumber
\end{eqnarray}
(see Appendix D, eq.(\ref{work-r})).
Therefore, the advanced expression can be replaced by the retarded one
plus functions of momentary positions of particles (see Appendix E, Table
2, left column, second line).

Now we consider the last terms in both eq.(\ref{G1}) and eq.(\ref{G2}).
Since the functions $t_1'(t,t_2)$ and $t_2'(t,t_1)$ are inverses, the sum
of these integrals can be written in the form either
\begin{equation}\label{ac1}
\int\limits_{t_1^{ret}(t)}^{t}d
t_1\left[G_2+\frac{1+V_1}{1-V_2}G_1\right]^{t_2=t_2'(t,t_1)}
\end{equation}
or
\begin{equation}\label{ac2}
\int\limits_{t_2^{ret}(t)}^{t}d
t_2\left[G_1+\frac{1-V_2}{1+V_1}G_2\right]^{t_1=t_1'(t,t_2)} .
\end{equation}
Both the expressions result the same function of the end points only:
\begin{equation}\label{1red}
-\left.\frac{\displaystyle 1}{\displaystyle 2(t-t_2)}
\right|_{t_2=t_2^{ret}(t)}^{t_2\to t}=-\lim_{t_2\to
t}\frac{e_1e_2}{2(t-t_2)}+\frac{e_1e_2}{2[t-t_2^{ret}(t)]}
\end{equation}
(see Appendix E, Table 1, left column, third line).

Summing up all the contributions (\ref{10ye}), (\ref{1ye}), (\ref{10blue})
where (\ref{1blue}) is taken into account, and (\ref{1red}) we obtain the
expression
\begin{eqnarray}\label{1rnb}
p^0_{int}&=&-\sum\limits_{b\ne
a}\int\limits_{-\infty}^tdt_a\gamma_a^{-1}F^0_{ba}(t_a,t_b^{ret}(t_a))\\
&-&\frac{e_1e_2}{q[t_1^{ret}(t),t]}
\frac{({\bf v}_2\cdot{\bf
v}_1)+V_2}{\left[1+V_1\right]\left[1+V_2\right]}-
\lim_{t_2\to t}\frac{e_1e_2}{t-t_2}\frac{V_2}{1+V_2}\nonumber
\end{eqnarray}

Now we take the double derivative in the form (\ref{dd2}) and add it
to $\partial G_2/\partial t_2$. Analogous calculations give
\begin{eqnarray}\label{2rnb}
p^0_{int}&=&-\sum\limits_{b\ne
a}\int\limits_{-\infty}^tdt_a\gamma_a^{-1}F^0_{ba}(t_a,t_b^{ret}(t_a))\\
&-&\frac{e_1e_2}{q[t,t_2^{ret}(t)]}
\frac{({\bf v}_1\cdot{\bf
v}_2)-V_1}{\left[1-V_1\right]\left[1-V_2\right]}+
\lim_{t_1\to t}\frac{e_1e_2}{t-t_1}\frac{V_1}{1-V_1}.\nonumber
\end{eqnarray}
(see Appendix E, Table 1, right column).

Having compared eq.(\ref{1rnb}) with (\ref{2rnb}) we are sure that the
calculations result the "immovable core" which describes the action of the
fields due to one charge on another, and "changeable shell" which
expresses the deformation of electromagnetic "clouds" of charged particles
due to mutual interaction. Only the immovable terms should be taken into
account in the total energy balance equation.

\subsection{Interference part of space components}

To calculate interference part $p_{int}^i$ of electromagnetic field
momentum $p_{em}^i$ we have to integrate the expression
\begin{equation}\label{Ti}
4\pi T_{int}^{0i}=f_{(1)}^{0j}f_{(2)j}^i + f_{(2)}^{0j}f_{(1)j}^i
\end{equation}
over three-dimensional hyperplane $y^0=t$. The electromagnetic field
components are given in Section 3.

According to the integration rules (\ref{int1}) and (\ref{int2}), first of
all we perform the angle integration. Then integrand (\ref{Ti})
looks as follows:
\begin{eqnarray} \label{Tif}
&&\frac{2}{e_1e_2}\int\limits_0^{2\pi}d\varphi\sqrt{-g}T_{int}^{0i}=\\
&=&{\cal A}_2^i\left[k_1^0\frac{\partial^2\lambda}{\partial t_1\partial
t_2}
+\frac{\partial\lambda}{\partial t_2}\right]
+{\cal C}_2^i\left[k_1^0\frac{\partial^3\lambda}{\partial t_1\partial
t_2^2}
+\frac{\partial^2\lambda}{\partial t_2^2}\right]
+{\cal B}_2^ik_1^0\frac{\partial^3\lambda}{\partial t_1^2\partial t_2}
+{\cal D}_2^ik_1^0\frac{\partial^4\lambda}{\partial t_1^2\partial t_2^2}
\nonumber\\
&+&{\cal A}_1^i\left[k_2^0\frac{\partial^2\lambda}{\partial t_1\partial
t_2}
+\frac{\partial\lambda}{\partial t_1}\right]+
{\cal C}_1^ik_2^0\frac{\partial^3\lambda}{\partial t_1\partial t_2^2}+
{\cal B}_1^i\left[k_2^0\frac{\partial^3\lambda}{\partial t_1^2\partial
t_2}
+\frac{\partial^2\lambda}{\partial t_1^2}\right]
+{\cal D}_1^ik_2^0\frac{\partial^4\lambda}{\partial t_1^2\partial t_2^2}
\nonumber\\
&+&{\cal
A}_0\left[\lambda(v_1^i+v_2^i)+k_2^0v_1^i\frac{\partial\lambda}{\partial
t_2}+k_1^0v_2^i\frac{\partial\lambda}{\partial t_1}
\right]
\nonumber\\
&+&{\cal C}_0\left[\lambda{\dot
v}_2^i+k_2^0v_1^i\frac{\partial^2\lambda}{\partial
t_2^2}+k_1^0{\dot v}_2^i\frac{\partial\lambda}{\partial t_1}
\right]
\nonumber\\
&+&{\cal B}_0\left[\lambda{\dot
v}_1^i+k_1^0v_2^i\frac{\partial^2\lambda}{\partial t_1^2}
+k_2^0{\dot v}_1^i\frac{\partial\lambda}{\partial
t_2}
\right]
\nonumber\\
&+&{\cal D}_0\left[k_2^0{\dot v}_1^i\frac{\partial^2\lambda}{\partial
t_2^2}+k_1^0{\dot v}_2^i\frac{\partial^2\lambda}{\partial t_1^2}
\right]\nonumber
\end{eqnarray}
where
\begin{equation}
\lambda = 1/2\left[(k_1^0-k_2^0)^2-q^2\right].
\end{equation}
Calligraphic letters ${\cal A, B, C}$ and ${\cal D}$ denote the
following integrals over $\varphi$:
\begin{eqnarray}\label{ABCD}
&&{\cal A}_b^i=\frac{1}{2\pi}\int\limits_0^{2\pi} d\varphi
K_b^i\frac{c_1c_2}{q(r_1)^2(r_2)^2},\quad
{\cal B}_b^i=\frac{1}{2\pi}\int\limits_0^{2\pi} d\varphi
K_b^i\frac{c_2}{qr_1(r_2)^2},\nonumber\\
&&{\cal C}_b^i=\frac{1}{2\pi}\int\limits_0^{2\pi} d\varphi
K_b^i\frac{c_1}{q(r_1)^2r_2},\quad
{\cal D}_b^i=\frac{1}{2\pi}\int\limits_0^{2\pi} d\varphi
K_b^i\frac{1}{qr_1r_2},
\end{eqnarray}
where $r_a$ and $c_a$ are given by eqs.(\ref{cr}). Functions
${\cal B}_0, {\cal C}_0$ and ${\cal D}_0$ are defined by
eqs.(\ref{BCD0}) and function ${\cal A}_0$ is
\begin{equation}
{\cal A}_0=\frac{1}{2\pi}\int\limits_0^{2\pi} d\varphi
\frac{c_1c_2}{q(r_1)^2(r_2)^2}.
\end{equation}

After some algebra one can rewrite the terms which involve ${\cal
A}_b^i, {\cal B}_b^i, {\cal C}_b^i$ and ${\cal D}_b^i$ (see the 1-st
and the 2-nd lines of eq.(\ref{Tif}) as follows:
\begin{eqnarray}\label{b-terms}
&&\frac{\partial}{\partial t_1}\left[
\Lambda_b\left({\cal B}_b^i-\frac{\partial {\cal D}_b^i}{\partial
t_2}\right)
\right]
+\frac{\partial}{\partial t_2}\left[
\Lambda_b\left({\cal C}_b^i-\frac{\partial {\cal D}_b^i}{\partial
t_1}\right)
\right]
+\frac{\partial^2 (\Lambda_b{\cal D}_b^i)}{\partial t_1\partial t_2}
\nonumber\\
&+&\Lambda_b\left({\cal A}_b^i-\frac{\partial {\cal B}_b^i}{\partial
t_1}
-\frac{\partial {\cal C}_b^i}{\partial t_2}+
\frac{\partial^2 {\cal D}_b^i}{\partial t_1\partial t_2}
\right).
\end{eqnarray}
Here
\begin{equation}\label{Lbd}
\Lambda_1=k_2^0\frac{\partial^2\lambda}{\partial t_1\partial t_2}+
\frac{\partial\lambda}{\partial t_1},\quad
\Lambda_2=k_1^0\frac{\partial^2\lambda}{\partial t_1\partial t_2}+
\frac{\partial\lambda}{\partial t_2}.
\end{equation}
Routine scrupulous calculations performed in Appendix B explain that
the "non-derivative tails" in eq.(\ref{b-terms}) are proportional to
three-velocities:
\begin{eqnarray}
{\cal A}_1^i-\frac{\partial {\cal B}_1^i}{\partial t_1}
-\frac{\partial {\cal C}_1^i}{\partial t_2}+
\frac{\partial^2 {\cal D}_1^i}{\partial t_1\partial t_2}&=&v_1^i(t_1)
\left({\cal B}_0-\frac{\partial {\cal D}_0}{\partial t_2}\right)\\
{\cal A}_2^i-\frac{\partial {\cal B}_2^i}{\partial t_1}
-\frac{\partial {\cal C}_2^i}{\partial t_2}+
\frac{\partial^2 {\cal D}_2^i}{\partial t_1\partial t_2}&=&v_2^i(t_2)
\left({\cal C}_0-\frac{\partial {\cal D}_0}{\partial
t_1}\right).\nonumber
\end{eqnarray}
We add them to the part of integrand (\ref{Tif}) which involve "zeroth"
functions ${\cal A}_0, {\cal B}_0, {\cal C}_0,$ and ${\cal D}_0$.
It is now straightforward (but tedious) matter to rewrite it as the
following sum of partial derivatives:
\begin{eqnarray}\label{v-terms}
&&\frac{\partial}{\partial t_1}\left[
\Lambda_2^i
\left({\cal B}_0-\frac{\partial {\cal D}_0}{\partial t_2}\right)
\right]+
\frac{\partial}{\partial t_2}\left[
\Lambda_2^i
\left({\cal C}_0-\frac{\partial {\cal D}_0}{\partial t_1}\right)
\right]+
\frac{\partial^2(\Lambda_2^i{\cal D}_0)}{\partial t_1\partial t_2}
-\frac{\partial (v_2^i\Lambda_2{\cal D}_0)}{\partial t_1}
\nonumber\\
&+&\frac{\partial}{\partial t_1}\left[
\Lambda_1^i
\left({\cal B}_0-\frac{\partial {\cal D}_0}{\partial t_2}\right)
\right]+
\frac{\partial}{\partial t_2}\left[
\Lambda_1^i
\left({\cal C}_0-\frac{\partial {\cal D}_0}{\partial t_1}\right)
\right]+
\frac{\partial^2(\Lambda_1^i{\cal D}_0)}{\partial t_1\partial t_2}
-\frac{\partial (v_1^i\Lambda_1{\cal D}_0)}{\partial t_2}
\nonumber\\
&+&\frac{\partial}{\partial t_1}\left[
\lambda\left(v_1^i+v_2^i\right)
\left({\cal B}_0-\frac{\partial {\cal D}_0}{\partial t_2}\right)
\right]+
\frac{\partial}{\partial t_2}\left[
\lambda\left(v_1^i+v_2^i\right)
\left({\cal C}_0-\frac{\partial {\cal D}_0}{\partial t_1}\right)
\right]\nonumber\\
&+&\frac{\partial^2}{\partial t_1\partial t_2}
\left[
\lambda\left(v_1^i+v_2^i\right){\cal D}_0
\right]
\end{eqnarray}
(We keep in mind the identity (\ref{A42})). Recall that $\Lambda_1,
\Lambda_2$ are given by eq.(\ref{Lbd}) and
\begin{equation}
\Lambda_1^i=v_1^ik_2^0\frac{\partial\lambda}{\partial t_2},\quad
\Lambda_2^i=v_2^ik_1^0\frac{\partial\lambda}{\partial t_1}.
\end{equation}

Therefore, the integrand (\ref{Tif}) also becomes the combinations of
partial derivatives with respect to time variables, namely the sum of
the expressions written in the first line of eq.(\ref{b-terms}) for both
$b=1$ and $b=2$, and eq.(\ref{v-terms}). Now we apply the integration
procedure developed in previous subsection.

Each double derivative involved in (\ref{Tif}) can be integrated
according to the rule either (\ref{int1}) or (\ref{int2}). There are five
terms of this type in this expression. This circumstance implies ten
possible ways of integrations. In Appendix E we study two of them in
detail (see Table 2 and Table 3).

Firstly we write all the double derivatives in the form (\ref{dd1}).
The integration gives
\begin{eqnarray}\label{pint1}
p^i_{int}&=&-\sum\limits_{b\ne
a}\int\limits_{-\infty}^tdt_a\gamma_a^{-1}F^i_{ba}(t_a,t_b^{ret}(t_a))\\
&+&\frac{e_1e_2}{q[t_1^{ret}(t),t]}\frac{\left[-1+({\bf v}_1\cdot{\bf
v}_2)\right]n_q^i[t_1^{ret}(t),t]}{\left[1+V_1\right]\left[1+V_2\right]}-
\frac{e_1e_2}{q[t_1^{ret}(t),t]}\frac{v_1^i[t_1^{ret}(t)]}{1+V_1}\nonumber\\
&+&\lim_{t_2\to
t}\frac{e_1e_2}{t-t_2}\frac{\left\{n_q^i[t_1^{ret}(t_2),t_2]-v_2^i(t_2)\right\}V_2}{1-(V_2)^2}
\nonumber
\end{eqnarray}
Secondly, we express all the mixed derivatives in the form (\ref{dd1}).
We obtain
\begin{eqnarray}\label{pint2}
p^i_{int}&=&-\sum\limits_{b\ne
a}\int\limits_{-\infty}^tdt_a\gamma_a^{-1}F^i_{ba}(t_a,t_b^{ret}(t_a))\\
&-&\frac{e_1e_2}{q[t,t_2^{ret}(t)]}\frac{\left[-1+({\bf v}_1\cdot{\bf
v}_2)\right]n_q^i[t,t_2^{ret}(t)]}{\left[1-V_1\right]\left[1-V_2\right]}-
\frac{e_1e_2}{q[t,t_2^{ret}(t)]}\frac{v_2^i[t_2^{ret}(t)]}{1-V_2}\nonumber\\
&+&\lim_{t_1\to
t}\frac{e_1e_2}{t-t_1}
\frac{\left\{n_q^i[t_1,t_2^{ret}(t_1)]+v_1^i(t_1)\right\}V_1}{1-(V_1)^2}
\nonumber
\end{eqnarray}
Comparing eq.(\ref{pint1}) with eq.(\ref{pint2}), we are sure that the form
of functions of momentary positions of particles heavily depend on the
method of integration. It reinforce our conviction that the changeable
"shell" expresses the deformation of electromagnetic "clouds" of "bare"
charges due to mutual interaction. Thus only the immovable "core", i.e. sum
of work done by Lorentz forces of point-like charges acting
on one another, possesses relevant physical sense.

\section{Conclusions}
\setcounter{equation}{0}

Inspection of the energy-momentum carried by the electromagnetic field of
two point-like charged particles reveals the essence of renormalization
procedure in classical electrodynamics. Volume integration of Maxwell
energy-momentum tensor density over three-dimensional hyperplane $y^0=t$
gives terms of two quite different types: (i) these depend on the state
of the particles' motion in the vicinity of the instant of observation $t$;
(ii) those depend on all previous time development of the sources. The
former involves diverging quantities while the latter contains finite
terms only. Structure of the quantities which are accumulated with time
does not depend on choice of integration three-surface while the form of
"instant" expression heavily depends on the way of integration.

"Instant" terms are permanently attached to the charges and are carried
along with them. By this we mean that a charged particle cannot be
separated from its bound electromagnetic "cloud" which has its own
4-momentum \cite{Teit}. This quantity together with 4-momentum of "bare"
charge
 constitute the {\it finite} 4-momentum of "dressed" charged particle.
(Note that the electromagnetic "clouds" of sources are deformed due to
mutual interaction.) All diverging quantities have thus disappeared into
the process of {\it energy-momentum renormalization}.

The terms which are accumulated with time lead to independent existence.
They constitute the radiative part of energy-momentum carried by
"two-particle" field. It consists of the integrals of individual Larmor
relativistic rates over corresponding world lines and the work done by
Lorentz forces of point-like charges acting on one another.

\begin{figure}[t]
\begin{center}
\epsfclipon
\epsfig{file=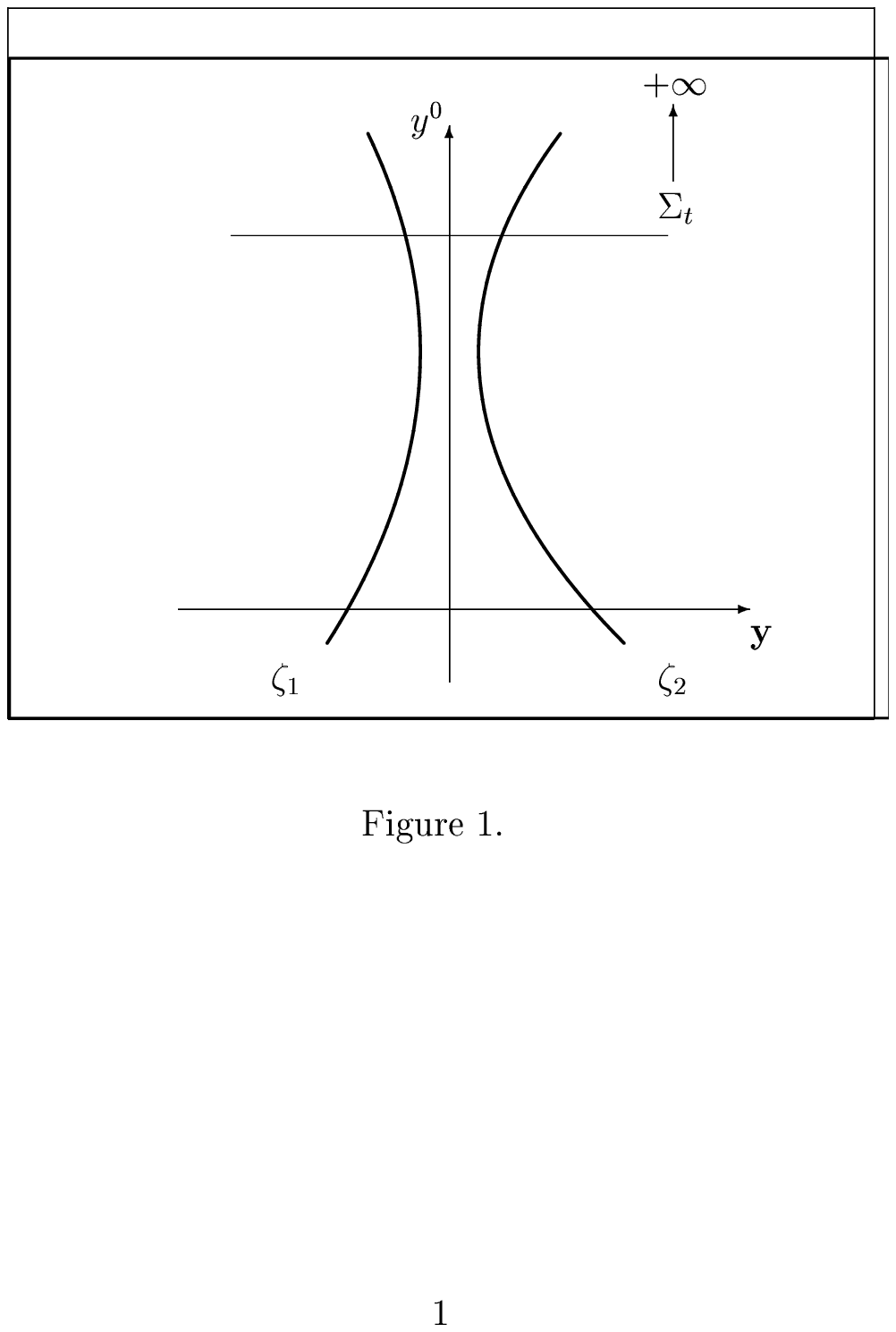,width=7cm}
\end{center}
\caption{\label{inf}
\small To restore time-reversal invariance we locate the observation
hyperplane $y^0=t$ in the distant future. We suppose that particles are
asymptotically free.
}
\end{figure}

The situation considered here, in which the radiation is propagating
outward, breaks the time-reversal invariance of Maxwell's theory. Choosing
the retarded solution of wave equation (\ref{w_e}) as the
physically-relevant solution, we adopt a specific time direction, when
{\it an interference of outgoing electromagnetic waves} leads to the
interaction between the sources. The interference are pictured in a fixed
{\it observation} hyperplane $\Sigma_t=\{y\in{\mathbb M}_4: y^0=t\}$. To
restore time-reversal invariance we take the limit $t\to +\infty$ and
suppose that particles are asymptotically free in the distant future. The
relation (\ref{d_wrk}) takes the form
\begin{equation}
\int\limits_{-\infty}^{+\infty}dt_a
\gamma_a^{-1}F_{ba}^\mu[t_a,t_b^{ret}(t_a)]
=\int\limits_{-\infty}^{+\infty}dt_b\gamma_b^{-1}F_{ab}^\mu
[t_a^{adv}(t_b),t_b].
\end{equation}
The work done by retarded Lorentz force of $b$-th charge over entire
world line of $a$-th one {\it is equal to} the work done by advanced
Lorentz force of $a$-th particle acting on $b$-th charge {\it backward} in
time! The sum of "retarded" works involved in the total energy-momentum of
our closed (two particles plus field) system
\begin{equation}
\int\limits_{-\infty}^{+\infty}dt_1
\gamma_1^{-1}F_{21}^\mu[t_1,t_2^{ret}(t_1)]+
\int\limits_{-\infty}^{+\infty}dt_2
\gamma_2^{-1}F_{12}^\mu[t_1^{ret}(t_2),t_2],
\end{equation}
may be replaced by the linear superposition
\begin{eqnarray}
&&\frac12\left[\int\limits_{-\infty}^{+\infty}dt_1 \gamma_1^{-1}
\left(F_{21}^\mu[t_1,t_2^{ret}(t_1)]+
F_{21}^\mu[t_1,t_2^{adv}(t_1)]
\right)\right.\nonumber\\
&+&\left.
\int\limits_{-\infty}^{+\infty}dt_2 \gamma_2^{-1}
\left(F_{12}^\mu[t_1^{ret}(t_2),t_2]+
F_{12}^\mu[t_1^{adv}(t_2),t_2]
\right)\right]
\end{eqnarray}
which restore time-reversal invariance. Indeed, the retarded Lorentz force
$F_{ba}^\mu[t_a,t_b^{ret}(t_a)]$ becomes the advanced one
$F_{ba}^\mu[t_a,t_b^{adv}(t_a)]$ (and vice versa) if the time direction
is reversed. But the retarded causality is still not violated. We consider
the interference of {\it outgoing} waves at distant future instead of a
picture in which the radiation is propagated inward.

The situation looks as that described by Wheeler and Feynman \cite{WF}
where the {\it absorber theory of radiation} is elaborated. The basic
assumption is that the fields which act on a given particle are
represented by one-half the retarded plus one-half the advanced
Li\'enard-Wiechert solutions of wave equations. To disappear "incoming"
radiation, the authors introduce a {\it perfect absorber} which cancels
the (acausal) advanced part of the fields acting on a given particle and
doubles the retarded one.

Our emphasis is on rigorous calculations and exact solutions based on
standard electrodynamics. It allows us to substitute the phenomenon of
interference of outgoing electromagnetic waves for acausal mechanism of
perfect absorbtion in {\it time-symmetric} action-at-a distance
electrodynamics. The interference of outgoing electromagnetic waves
(retarded Li\'enard-Wiechert solutions) ensures the action of the field of
one source on another (mutual interaction).

\section*{Acknowledgments}

The author would like to thank Professor V.Tre\-tyak  and
Dr. A.Du\-vi\-ryak for helpful discussions and comments.

\subsubsection{Angle integration of $T_{\rm int}^{00}$}
\renewcommand{\theequation}{\Alph{subsubsection}.\arabic{equation}}
\setcounter{equation}{0}
In this Appendix we perform the integration over $\varphi$ of
"double zeroth" component (\ref{JT00}) of the Maxwell energy-momentum
tensor density. Angle-dependent terms involved in energy density have the
form
\begin{eqnarray}\label{abcd}
{\cal A}_0&=&\frac{1}{2\pi}\int\limits_{0}^{2\pi}d\varphi
\frac{c_1c_2}{q(r_1)^2(r_2)^2},
\qquad
{\cal B}_0=\frac{1}{2\pi}\int\limits_{0}^{2\pi}d\varphi
\frac{c_2}{qr_1(r_2)^2},
\\
{\cal C}_0&=&\frac{1}{2\pi}\int\limits_{0}^{2\pi}d\varphi
\frac{c_1}{q(r_1)^2r_2},
\qquad\,\,\,\,\,\,\,
{\cal D}_0=\frac{1}{2\pi}\int\limits_{0}^{2\pi}d\varphi
\frac{1}{qr_1r_2}
\nonumber
\end{eqnarray}
where the retarded distances are
\begin{equation}\label{ra}
r_a=r_a^0-r_a^1\sin\varphi-r_a^2\cos\varphi .
\end{equation}
The other scalars we use are:
\begin{equation}\label{ca}
c_a=-c_a^0+c_a^1\sin\varphi+c_a^2\cos\varphi
\end{equation}
Here
\begin{eqnarray}\label{rcomp}
r_a^0&=&k_a^0-({\bf v}_a\cdot{\bf n}_q)k_a^3, \quad
r_a^1=\omega_{1j}v_a^jh,\quad r_a^2=\omega_{2j}v_a^jh\\
c_a^0&=&-\left(\gamma_a^{-2}+({\bf \dot{v}}_a\cdot{\bf n}_q)k_a^3\right),
\quad c_a^1=\omega_{1j}\dot{v}_a^jh,\quad c_a^2=\omega_{2j}\dot{v}_a^jh.
\label{ccomp}
\end{eqnarray}

It is convenient to introduce three-dimensional manifold $\mathbb Q$ with
space-favouring metric $g_{\alpha\beta}={\rm diag}(-1,1,1)$. For $\mathbb Q$
we put the tangent bundle $T\mathbb Q$ being the disjoint union of all
tangent spaces $T_x\mathbb Q$. A tangent vector with foot point
$a\in \mathbb Q$ is simply a pair $(a,{\bf r})$ with
$$
{\bf r}=r^\alpha{\bf e}_\alpha\in{\mathbb R}^3,
$$
where ${\bf e}_\alpha:=\partial/\partial x^\alpha, \alpha=0,1,2$, is the
standard basis of ${\mathbb R}^3$. We define also cotangent bundle
$T^*\mathbb Q$ being the disjoint union of $T^*_x\mathbb Q$. An one-form
with foot point $a\in \mathbb Q$ is a pair $(a,{\hat {\bf r}})$ with
$$
{\hat {\bf r}}=r_\beta{\hat {\bf e}}^\beta\in{\mathbb R}^3,
$$
where ${\hat {\bf e}}^\beta, \beta=0,1,2$, constitute dual basis
${\hat {\bf e}}^\beta({\bf e}_\alpha)=\delta_\alpha^\beta$. We shall use
$g_{\alpha\beta}={\rm diag}(-1,1,1)$ and its inverse $g^{\alpha\beta}={\rm
diag}(-1,1,1)$ to lower and raise indices, respectively.

For each differential manifold one can define the canonical pairing
\begin{eqnarray}
\langle ,\rangle &:&T^*{\mathbb Q}\times_{\mathbb Q}T{\mathbb Q}\to
{\mathbb R}\\
&&\langle {\hat {\bf r}}_1,{\bf r}_2\rangle \mapsto r_{1,\gamma}r_2^\gamma
\nonumber
\end{eqnarray}
where both one-form ${\hat {\bf r}}_1$ and vector ${\bf r}_2$ are of the
same foot point. We introduce also the scalar product
\begin{eqnarray}
(\cdot)&:&T{\mathbb Q}\times_{\mathbb Q}T{\mathbb Q}\to \mathbb R\\
&&({\bf r}_1\cdot{\bf r}_2)\mapsto g_{\alpha\beta}r_1^\alpha r_2^\beta
\nonumber
\end{eqnarray}
which is connected with canonical pairing by the operation of raising
indices. And finally, we shall need the norm of vector
${\bf r}\in T{\mathbb Q}$
\begin{eqnarray}
\|{\bf r}\|&=&\sqrt{-g_{\alpha\beta}r^\alpha r^\beta }\\
&=&\sqrt{(r^0)^2-(r^1)^2-(r^2)^2}\nonumber
\end{eqnarray}
So, $a$-th retarded distance $r_a$ becomes the scalar product of
the vector with components (\ref{rcomp}) and the null-vector
${\bf n}_\varphi:=(1,\sin\varphi,\cos\varphi)$ taken with opposite sign.

To go further we express a term of type $({\bf b}\cdot{\bf
n}_\varphi)/r_1r_2$ as follows
\begin{equation} \label{basic}
\frac{b}{r_1r_2}=\frac{-A-C|{\bf r}_1|\cos
(\varphi+\beta_1)}{r_1^0-|{\bf r}_1|\sin(\varphi+\beta_1)}
+
\frac{-B+C|{\bf r}_2|\cos
(\varphi+\beta_2)}{r_2^0-|{\bf r}_2|\sin(\varphi+\beta_2)}
\end{equation}
where scalar $b$ denotes the product $({\bf b}\cdot {\bf n}_\varphi)$.
The $a$-th phase $\beta_a$ is determined by the relations
\begin{equation}
\cos\beta_a = r_a^1/|{\bf r}_a|,\quad \sin\beta_a = r_a^2/|{\bf r}_a|,
\end{equation}
where
\begin{eqnarray}
|{\bf r}_a|&=&\sqrt{(r_a^1)^2+(r_a^2)^2}\nonumber\\
&=&h\sqrt{{\bf v}_a^2-({\bf v}_a\cdot{\bf n}_q)^2}.
\end{eqnarray}
The coefficients $A, B$ and $C$ are the solutions of the following system
of algebraic equations
\begin{equation}\label{l}
\begin{array}{ccc}
(C & A & B)\\
&&\\
 &  &
\end{array}
\left(
\begin{array}{ccc}
L_0 & L_1 & L_2\\[2pt]
r_{2,0}&r_{2,1}&r_{2,2}\\[4pt]
r_{1,0}&r_{1,1}&r_{1,2}
\end{array}
\right)=
\begin{array}{ccc}
(b_0 & b_1 & b_2)\\
&&\\
 &  &
\end{array}
\end{equation}
where $L_\alpha=g_{\alpha\beta}L^\beta$ and $L^\beta$ is $^\beta$-th
component of the vector
\begin{equation}
L=\left(
\begin{array}{ccc}
{\bf e}_0 & {\bf e}_1 & {\bf e}_2\\[2pt]
r_{2,0}&r_{2,1}&r_{2,2}\\[4pt]
r_{1,0}&r_{1,1}&r_{1,2}
\end{array}
\right)
\end{equation}
This can be written more compactly
$$
L^\alpha=\varepsilon^{\alpha\beta\gamma}r_{2,\beta}r_{1,\gamma}
$$
by use of the Ricci symbol in three dimensions:
\begin{equation}
\varepsilon^{\alpha\beta\gamma}=\left\{
\begin{array}{l}
+1 \,\,\,{\rm when}\,\,\alpha\beta\gamma\,\, {\rm is}\,\,{\rm an}\,\,{\rm
even}\,\,{\rm permutation}\,\,{\rm of}\,\, 0,1,2 \\[2pt]
-1 \,\,\,{\rm when}\,\,\alpha\beta\gamma\,\, {\rm is}\,\,{\rm an}\,\,{\rm
odd} \,\, {\rm permutation}\,\,{\rm of}\,\, 0,1,2 \\[2pt]
\,0\,\,\, {\rm otherwise}
\end{array}
\right.
\end{equation}

In solving the problem (\ref{basic}) one is soon led into rather complex
expression. Great simplification arise, however, when one uses the binary
operation of vector product which is defined as follows:
\begin{eqnarray}\label{v-pr}
[\,\times\,]&:&T^*{\mathbb Q}\times_{\mathbb Q}T^*{\mathbb Q}\to T{\mathbb
Q}\\
&&[{\hat {\bf r}}_1\times{\hat {\bf r}}_2]\mapsto
\varepsilon^{\alpha\beta\gamma}r_{1,\beta}r_{2,\gamma} \nonumber
\end{eqnarray}
So, the determinant $D$ of $3\times 3$ matrix in eq.(\ref{l}) becomes the
square of vector product of one-forms ${\hat {\bf r}}_1$ and
${\hat {\bf r}}_2$ given by eqs.(\ref{rcomp}):
\begin{eqnarray}\label{D}
D&=&\langle {\bf\hat L},{\bf L}\rangle =({\bf L})^2
\\
&=&[{\hat {\bf r}}_2\times{\hat {\bf r}}_1]^2
\nonumber\\
&=&-\langle {\bf\hat r}_2,{\bf r}_2\rangle \langle {\bf\hat r}_1,{\bf
r}_1\rangle +
\langle {\bf\hat r}_2, {\bf r}_1\rangle ^2
\nonumber\\
&=&-({\bf r}_2\cdot{\bf r}_2)({\bf r}_1\cdot{\bf r}_1)+({\bf r}_2\cdot{\bf
r}_1)^2
\nonumber
\end{eqnarray}
Having solved the system of linear equations (\ref{l}) we obtain
\begin{eqnarray}\label{ABC}
A&=&\frac{({\bf b}\cdot[{\bf\hat r}_1\times{\bf\hat L}])}{D}
\\
&=&-\frac{1}{D}\left\{\phantom{\frac11\!\!\!\!}
({\bf b}\cdot{\bf r}_2)({\bf r}_1\cdot{\bf r}_1)
-({\bf b}\cdot{\bf r}_1)({\bf r}_1\cdot{\bf r}_2)
\right\}
\nonumber\\
B&=&-\frac{({\bf b}\cdot[{\bf\hat r}_2\times{\bf\hat L}])}{D}
\\
&=&\frac{1}{D}\left\{\phantom{\frac11\!\!\!\!}
({\bf b}\cdot{\bf r}_2)({\bf r}_2\cdot{\bf r}_1)
-({\bf b}\cdot{\bf r}_1)({\bf r}_2\cdot{\bf r}_2)
\right\}
\nonumber\\
C&=&\frac{({\bf b}\cdot{\bf L})}{D}.
\end{eqnarray}
The expression of type (\ref{basic}) can be integrated over $\varphi$ via
the relations
\begin{eqnarray} \label{fi}
\int_0^{2\pi}d\varphi\frac{1}{1-a\sin\varphi}&=&\frac{2\pi}{\sqrt{1-a^2}},\qquad
\int_0^{2\pi}d\varphi\frac{\cos\varphi}{1-a\sin\varphi}=0\\
&&0\le a<1.\nonumber
\end{eqnarray}
Having integrated (\ref{basic}) we obtain
\begin{equation}
\int_0^{2\pi}d\varphi\frac{b}{r_1r_2}=-2\pi\left(\frac{A}{\|{\bf r}_1\|}+
\frac{B}{\|{\bf r}_2\|}\right).
\end{equation}

Having considered the simplest case of integral ${\cal D}_0$ (see
eqs.(\ref{ABCD})), we put the one-form ${\bf\hat b}=(-1,0,0)$. We obtain
\begin{equation}\label{D_0}
{\cal D}_0=-\frac{1}{q}\left(\frac{A_0}{\|{\bf r}_1\|}+
\frac{B_0}{\|{\bf r}_2\|}\right),
\end{equation}
where
\begin{eqnarray}\label{yel-cf}
A_0&=&\frac{\varepsilon^{0\alpha\beta}r_{1,\alpha}L_\beta}{D}
\\
&=&-\frac{r_2^0({\bf r}_1\cdot{\bf r}_1)
-r_1^0({\bf r}_2\cdot{\bf r}_1)}{D}
\nonumber\\
B_0&=&-\frac{\varepsilon^{0\alpha\beta}r_{2,\alpha}L_\beta}{D}
\nonumber\\
&=&\frac{r_2^0({\bf r}_2\cdot{\bf r}_1)
-r_1^0({\bf r}_2\cdot{\bf r}_2)}{D}
\nonumber\\
C_0&=&\frac{L^0}{D}
\nonumber
\end{eqnarray}

To calculate ${\cal B}_0$ we rewrite the integrand as follows
\begin{eqnarray} \label{green}
\frac{c_2}{r_1(r_2)^2}&=&\frac{c_2}{r_1r_2}\frac{1}{r_2}
\nonumber\\
&=&\left(\frac{-A_2-C_2|{\bf r}_1|\cos
(\varphi+\beta_1)}{r_1}+\frac{-B_2+C_2|{\bf r}_2|\cos
(\varphi+\beta_2)}{r_2}\right)\frac{1}{r_2}
\nonumber\\
&=&\frac{-A_2'-C_2'|{\bf r}_1|\cos
(\varphi+\beta_1)}{r_1^0-|{\bf r}_1|\sin(\varphi+\beta_1)}
+
\frac{-B_2'+C_2'|{\bf r}_2|\cos
(\varphi+\beta_2)}{r_2^0-|{\bf r}_2|\sin(\varphi+\beta_2)}\nonumber\\[5pt]
&+&\frac{-B_2+C_2|{\bf r}_2|\cos
(\varphi+\beta_2)}{\left[r_2^0-|{\bf r}_2|\sin(\varphi+\beta_2)\right]^2},
\end{eqnarray}
where
\begin{eqnarray}\label{gre-cf}
A_2&=&\frac{({\bf c}_2\cdot[{\bf\hat r}_1\times{\bf\hat L}])}{D},\quad
B_2=-\frac{({\bf c}_2\cdot[{\bf\hat r}_2\times{\bf\hat L}])}{D},\quad
C_2=\frac{({\bf c}_2\cdot{\bf L})}{D};
\\
A_2'&=&-A_2A_0+C_2C_0\|{\bf r}_1\|^2,\quad
B_2'=-A_2B_0+C_2C_0({\bf r}_2\cdot{\bf r}_1),
\nonumber\\
C_2'&=&-A_2C_0-C_2A_0
\label{gre}
\end{eqnarray}
Another relevant integration rules are
\begin{eqnarray}\label{fi^2}
\int_0^{2\pi}d\varphi\frac{1}{(1-a\sin\varphi)^2}&=&\frac{2\pi}{(1-a^2)^{3/2}},\qquad
\int_0^{2\pi}d\varphi\frac{\cos\varphi}{(1-a\sin\varphi)^2}=0\\
&&0\le a<1.\nonumber
\end{eqnarray}
Combining these results together with the relations (\ref{fi}) for
integral of expression (\ref{green}) over $\varphi$ gives
\begin{equation}\label{B_0}
{\cal B}_0=-\frac{1}{q}\left(\frac{A_2'}{\|{\bf r}_1\|}+
\frac{B_2'}{\|{\bf r}_2\|}+\frac{B_2r_2^0}{\|{\bf r}_2\|^3}\right).
\end{equation}

If one interchanges the indices "first" and "second" in the above
expression (\ref{green}), they obtain
\begin{equation}\label{C_0}
{\cal C}_0=-\frac{1}{q}\left(\frac{A_1'}{\|{\bf r}_1\|}+
\frac{B_1'}{\|{\bf r}_2\|}+\frac{A_1r_1^0}{\|{\bf r}_1\|^3}\right),
\end{equation}
where
\begin{eqnarray}\label{blu-cf}
A_1&=&\frac{({\bf c}_1\cdot[{\bf\hat r}_1\times{\bf\hat L}])}{D},\quad
B_1=-\frac{({\bf c}_1\cdot[{\bf\hat r}_2\times{\bf\hat L}])}{D},\quad
C_1=\frac{({\bf c}_1\cdot{\bf L})}{D};
\\
A_1'&=&-B_1A_0+C_1C_0({\bf r}_2\cdot{\bf r}_1),\quad
B_1'=-B_1B_0+C_1C_0\|{\bf r}_2\|^2,
\nonumber\\
C_1'&=&-B_1C_0-C_1B_0.
\label{blu}
\end{eqnarray}

We now turn to the calculation of ${\cal A}_0$. Transformation of the
integrand scales as $(r_1r_2)^{-2}$ proceeds with the help of
eq.(\ref{basic}), using identity
\begin{equation}
\frac{c_1c_2}{(r_1r_2)^2}=\frac{c_1}{r_1r_2}\frac{c_2}{r_1r_2}.
\end{equation}
The calculation is straightforward, although it involves a fair amount of
algebra. Finally we obtain
\begin{eqnarray} \label{red}
\frac{c_1}{r_1r_2}\frac{c_2}{r_1r_2}&=&
\frac{I_1+I_0|{\bf r}_1|\cos (\varphi+\beta_1)}{(r_1)^2}+
\frac{J_1-J_0|{\bf r}_2|\cos (\varphi+\beta_2)}{(r_2)^2}
\\
&+&
\frac{-A_{12}-C_{12}|{\bf r}_1|\cos (\varphi+\beta_1)}{r_1}
+
\frac{-B_{12}+C_{12}|{\bf r}_2|\cos (\varphi+\beta_2)}{r_2}
\nonumber
\end{eqnarray}
where
\begin{eqnarray}\label{red-cf}
I_1&=&A_1A_2-C_1C_2\|{\bf r}_1\|^2,\qquad I_0=A_1C_2+A_2C_1,
\\
J_1&=&B_1B_2-C_1C_2\|{\bf r}_2\|^2,\qquad J_0=B_1C_2+B_2C_1,
\nonumber\\
A_{12}&=&\left[A_1B_2+B_1A_2-2C_1C_2({\bf r}_1\cdot{\bf r}_2)\right]A_0
-J_0C_0\|{\bf r}_1\|^2-I_0C_0({\bf r}_1\cdot{\bf r}_2),
\nonumber\\
B_{12}&=&\left[A_1B_2+B_1A_2-2C_1C_2({\bf r}_1\cdot{\bf r}_2)\right]B_0
-J_0C_0({\bf r}_1\cdot{\bf r}_2)-I_0C_0\|{\bf r}_2\|^2,
\nonumber\\
C_{12}&=&\left[A_1B_2+B_1A_2-2C_1C_2({\bf r}_1\cdot{\bf r}_2)\right]C_0
+I_0B_0+J_0A_0.
\end{eqnarray}
Using integration rules (\ref{fi}) and (\ref{fi^2}), we perform
the integration of (\ref{red}) over the angle variable $\varphi$:
\begin{equation}\label{A_0}
{\cal A}_0=-\frac{1}{q}\left(\frac{A_{12}}{\|{\bf
r}_1\|}+\frac{B_{12}}{\|{\bf r}_2\|}-\frac{I_1r_1^0}{\|{\bf
r}_1\|^3}-\frac{J_1r_2^0}{\|{\bf
r}_2\|^3}\right).
\end{equation}

All the coefficients involved in resulting expressions (\ref{D_0}),
(\ref{B_0}), (\ref{C_0}), and (\ref{A_0}) should be rewritten in terms of
three-dimensional vectors which denote particles' positions, velocities and
accelerations. Substituting components (\ref{rcomp}) and (\ref{ccomp})
into expressions (\ref{yel-cf}), (\ref{gre-cf}) and (\ref{blu-cf}) returns
the root coefficients $A_i, B_i$ and $C_i, i=0,1,2$:
\begin{eqnarray}\label{A_yl}
A_0&=&-\frac{([{\bf n}_q\times {\bf
v}_1]\cdot [{\bf n}_q\times {\bf l}])}{\Delta},\\
B_0&=&\frac{([{\bf n}_q\times {\bf
v}_2]\cdot [{\bf n}_q\times {\bf l}])}{\Delta},
\nonumber\\
C_0&=&\frac{({\bf n}_q\cdot [{\bf v}_2\times
{\bf v}_1])}{\Delta},\nonumber
\end{eqnarray}
where
\begin{equation}\label{delta}
\Delta = [{\bf n}_q\times {\bf l}]^2-h^2({\bf n}_q[{\bf v}_2\times {\bf
v}_1])^2,\quad {\bf l}=r_2^0{\bf v}_1-r_1^0{\bf v}_2;
\end{equation}
\begin{eqnarray}\label{A_bl}
A_1&=&\Delta^{-1}\left\{\phantom{\!\!\!\!\frac11}
c_1^0\left(
\left[{\bf n}_q\times{\bf v}_1\right]\cdot
\left[{\bf n}_q\times{\bf l}\right]
\right) -
r_1^0\left(
\left[{\bf n}_q\times{\bf\dot v}_1\right]\cdot
\left[{\bf n}_q\times{\bf l}\right]
\right)\right.\nonumber\\
&+&\left. h^2\left({\bf n}_q\cdot\left[{\bf\dot v}_1\times{\bf
v}_1\right]\right)
\left({\bf n}_q\cdot\left[{\bf v}_1\times{\bf
v}_2\right]\right)
\phantom{\!\!\!\!\!\!\frac11}\right\},\\
B_1&=&-\Delta^{-1}\left\{\phantom{\!\!\!\!\frac11}
c_1^0\left(
\left[{\bf n}_q\times{\bf v}_2\right]\cdot
\left[{\bf n}_q\times{\bf l}\right]
\right) -
r_2^0\left(
\left[{\bf n}_q\times{\bf\dot v}_1\right]\cdot
\left[{\bf n}_q\times{\bf l}\right]
\right)\right.\nonumber\\
&+&\left. h^2\left({\bf n}_q\cdot\left[{\bf\dot v}_1\times{\bf
v}_2\right]\right)
\left({\bf n}_q\cdot\left[{\bf v}_1\times{\bf
v}_2\right]\right)
\phantom{\!\!\!\!\!\!\frac11}\right\},\nonumber\\
C_1&=&\Delta^{-1}\left\{\phantom{\!\!\!\!\!\!\frac11}
-c_1^0\left({\bf n}_q\cdot\left[{\bf v}_2\times{\bf v}_1\right]\right) +
\left({\bf n}_q\cdot\left[{\bf\dot v}_1\times{\bf l}\right]\right)
\right\},\nonumber
\end{eqnarray}

\begin{eqnarray}\label{A_gr}
A_2&=&\Delta^{-1}\left\{\phantom{\!\!\!\!\frac11}
c_2^0\left(
\left[{\bf n}_q\times{\bf v}_1\right]\cdot
\left[{\bf n}_q\times{\bf l}\right]
\right) -
r_1^0\left(
\left[{\bf n}_q\times{\bf\dot v}_2\right]\cdot
\left[{\bf n}_q\times{\bf l}\right]
\right)\right.\nonumber\\
&+&\left. h^2\left({\bf n}_q\cdot\left[{\bf\dot v}_2\times{\bf
v}_1\right]\right)
\left({\bf n}_q\cdot\left[{\bf v}_1\times{\bf
v}_2\right]\right)
\phantom{\!\!\!\!\frac11}\right\},\\
B_2&=&-\Delta^{-1}\left\{\phantom{\!\!\!\!\frac11}
c_2^0\left(
\left[{\bf n}_q\times{\bf v}_2\right]\cdot
\left[{\bf n}_q\times{\bf l}\right]
\right) -
r_2^0\left(
\left[{\bf n}_q\times{\bf\dot v}_2\right]\cdot
\left[{\bf n}_q\times{\bf l}\right]
\right)\right.\nonumber\\
&+&\left. h^2\left({\bf n}_q\cdot\left[{\bf\dot v}_2\times{\bf
v}_2\right]\right)
\left({\bf n}_q\cdot\left[{\bf v}_1\times{\bf
v}_2\right]\right)
\phantom{\!\!\!\!\frac11}\right\},\nonumber\\
C_2&=&\Delta^{-1}\left\{\phantom{\!\!\!\!\!\!\frac11}
-c_2^0\left({\bf n}_q\cdot\left[{\bf v}_2\times{\bf v}_1\right]\right) +
\left({\bf n}_q\cdot\left[{\bf\dot v}_2\times{\bf l}\right]\right)
\right\}.\nonumber
\end{eqnarray}

A complex calculation performed with the help of software system "Maple 8"
confirms the key identity
\begin{equation}\label{A42}
{\cal A}_0-\frac{\partial {\cal B}_0}{\partial t_1}
-\frac{\partial {\cal C}_0}{\partial t_2}+
\frac{\partial^2 {\cal D}_0}{\partial t_1\partial t_2}=0.
\end{equation}
It allows us to rewrite the integral of "double zeroth" component of the
Maxwell energy-momentum tensor density over $\varphi$ as the sum
(\ref{fiT00}) of partial derivatives in time variables.

It is worth noting that all the coefficients (\ref{A_yl})-(\ref{A_gr})
and, therefore, expressions (\ref{D_0}), (\ref{B_0}), (\ref{C_0}) and
(\ref{A_0}) depend on $h^2$, i.e. on the square of the radius of the
circle $C(O,h)=S_1\cap S_2$ (see Figs.\ref{intf},\ref{k_k}). One can
express functions ${\cal A}_0$, ${\cal B}_0$, ${\cal C}_0$ and ${\cal
D}_0$ in form of expansions in powers of $h^2$. (To simplify the
calculations as much as possible we can rewrite the {\it integrands} of
(\ref{abcd})
as expansions in power $h$ and then {\it integrate over} $\varphi$.)
Putting $h^2\to 0$ we tend to convex neighbourhood of the end points,
either $S$ or $N$ (see Figs.\ref{ra_cr}, \ref{nc-cr}). The identity
(\ref{A42}) is also valid in the immediate vicinity of the end points.
(Differentiating functions ${\cal B}_0$, ${\cal C}_0$ and ${\cal D}_0$ in
time variables we must keep in mind that $\partial h^2/\partial t_a$ does
not vanish even if $h^2\to 0$.)

\subsubsection{Angle integration of $T_{\rm int}^{0i}$}
\renewcommand{\theequation}{\Alph{subsubsection}.\arabic{equation}}
\setcounter{equation}{0}

To express the integral of $T_{\rm int}^{0i}$ over $\varphi$ as a
combination of partial derivatives in time variables we have to calculate
the following "tails":
\begin{equation} \label{tail_i}
{\cal A}_b^i-\frac{\partial {\cal B}_b^i}{\partial t_1}
-\frac{\partial {\cal C}_b^i}{\partial t_2}+
\frac{\partial^2 {\cal D}_b^i}{\partial t_1\partial t_2}
\end{equation}
(see eq.(\ref{b-terms}). By calligraphic letters we denote the integrals
over angle variable:
\begin{eqnarray}\label{ABCD-i}
{\cal A}_b^i&=&\frac{1}{2\pi}\int\limits_{0}^{2\pi}d\varphi
K_b^i\frac{c_1c_2}{q(r_1)^2(r_2)^2},
\qquad
{\cal B}_b^i=\frac{1}{2\pi}\int\limits_{0}^{2\pi}d\varphi
K_b^i\frac{c_2}{qr_1(r_2)^2},
\\
{\cal C}_b^i&=&\frac{1}{2\pi}\int\limits_{0}^{2\pi}d\varphi
K_b^i\frac{c_1}{q(r_1)^2r_2},
\qquad\,\,\,\,\,\,\,
{\cal D}_b^i=\frac{1}{2\pi}\int\limits_{0}^{2\pi}d\varphi
\frac{K_b^i}{qr_1r_2}
\nonumber
\end{eqnarray}
where
\begin{equation}
K_b^i=k_b^3n_q^i+h\omega_1^i\sin\varphi	+h\omega_2^i\cos\varphi	.
\end{equation}
The integration can be handled via the relation (\ref{basic}). The
simplest term ${\cal D}_b^i$ becomes:
\begin{equation}
{\cal D}_b^i=-\frac{1}{q}\left(\frac{A_b^i}{\|{\bf r}_1\|}+
\frac{B_b^i}{\|{\bf r}_2\|}\right).
\end{equation}
Here
\begin{eqnarray}
A_b^i=k_b^3n_q^iA_0&-&\frac{1}{\Delta}
\left\{\phantom{\!\!\!\!\frac11}\left[v_2^i-n_q^i({\bf n}_q{\bf
v}_2)\right]
({\bf r}_1\cdot{\bf r}_1)\right.\nonumber
\\
&-&\left.\left[v_1^i-n_q^i({\bf n}_q{\bf v}_1)\right]
({\bf r}_1\cdot{\bf r}_2)\phantom{\!\!\!\!\frac11}\right\},
\\
B_b^i=k_b^3n_q^iB_0&+&\frac{1}{\Delta}
\left\{\phantom{\!\!\!\!\frac11}\left[v_2^i-n_q^i({\bf n}_q{\bf
v}_2)\right]
({\bf r}_2\cdot{\bf r}_1)\right.\nonumber
\\
&-&\left.\left[v_1^i-n_q^i({\bf n}_q{\bf v}_1)\right]
({\bf r}_2\cdot{\bf r}_2)\phantom{\!\!\!\!\frac11}\right\}
\nonumber
\end{eqnarray}
where $A_0$, $B_0$ and $\Delta$ are defined by
eqs.(\ref{A_yl}) and (\ref{delta}). But we find out the expressions
(\ref{tail_i}) in another way.

To simplify the calculations as much as possible we express the integrands
of eqs.(\ref{ABCD-i}) in form of expansions in powers of $h$. Thanks to
exponential operator
\begin{equation}\label{Y}
Y:=\exp\left[-\sum_ahv_a^j(\omega_{j1}\sin\varphi+\omega_{j2}\cos\varphi)
\frac{d}{dr_a^0}
\right]
\end{equation}
we remove harmonic functions from denominators and then integrate
over $\varphi$. In fact, we deal with the flow of the vector field in
between the square brackets of eq.(\ref{Y}). It maps an open neighbourhood
of end points either $S$ or $N$ to an open vicinity of another point of
integral curve of this vector field \cite{KMS}. It is sufficient to
compute "tails" (\ref{tail_i}) at the end points where
$h^2=0$ (see Figs.\ref{ra_cr} and \ref{nc-cr}).

At these end points the term ${\cal A}_b^i$ is as follows:
\begin{eqnarray}\label{A^ih}
{\cal A}_b^i&=&\int\limits_{0}^{2\pi}d\varphi
\left.K_b^i\frac{c_1c_2}{q(r_1)^2(r_2)^2}\right|_{h^2=0}
\\
&=&k_b^3n_q^i\frac{c_1^0c_2^0}{q(r_1^0)^2(r_2^0)^2}
\nonumber\\[5pt]
&=&\left.k_b^3n_q^i{\cal A}_0\right|_{h^2=0}.
\nonumber
\end{eqnarray}
Since derivatives $\partial h^2/\partial t_a, a=1,2,$ do not vanish whenever
$h^2=0$, we should expand ${\cal C}_b^i$ and ${\cal B}_b^i$ up to the first
order of this small parameter:
\begin{eqnarray}
{\cal B}_b^i&=&k_b^3n_q^i{\cal B}_0
\\
&+&\frac{h^2}{2qr_1^0(r_2^0)^2}\left[
-c_2^0\left(2\frac{v_2^i-n_q^i({\bf n}_q{\bf v}_2)}{r_2^0}+
\frac{v_1^i-n_q^i({\bf n}_q{\bf v}_1)}{r_1^0}\right)
+{\dot v}_2^i-n_q^i({\bf n}_q{\dot{\bf v}}_2)
\right] \nonumber\\[1ex]
&+&O(h^2)
\nonumber\\[1ex]
{\cal C}_b^i&=&k_b^3n_q^i{\cal C}_0
\\
&+&\frac{h^2}{2q(r_1^0)^2r_2^0}\left[
-c_1^0\left(\frac{v_2^i-n_q^i({\bf n}_q{\bf v}_2)}{r_2^0}+
2\frac{v_1^i-n_q^i({\bf n}_q{\bf v}_1)}{r_1^0}\right)
+{\dot v}_1^i-n_q^i({\bf n}_q{\dot{\bf v}}_1)
\right]
\nonumber\\[1ex]
&+&O(h^2)
\nonumber
\end{eqnarray}
Symbols ${\cal B}_0$ and ${\cal C}_0$ denote the expansions of
corresponding integrals (\ref{ABCD}) in powers of $h^2$:
\begin{eqnarray}
{\cal B}_0&=&\frac{-c_2^0}{qr_1^0(r_2^0)^2}
\\
&+&\frac{-c_2^0}{2qr_1^0(r_2^0)^2}\left[
3\frac{[{\bf n}_q{\bf v}_2]^2}{(r_2^0)^2}+
2\frac{([{\bf n}_q{\bf v}_1][{\bf n}_q{\bf v}_2])}{r_1^0r_2^0}
+\frac{[{\bf n}_q{\bf v}_1]^2}{(r_1^0)^2}
\right]h^2
\nonumber\\
&+&\frac{1}{2qr_1^0(r_2^0)^2}\left[
2\frac{([{\bf n}_q{\dot {\bf v}}_2][{\bf n}_q{\bf v}_2])}{r_2^0}+
\frac{([{\bf n}_q{\dot {\bf v}}_2][{\bf n}_q{\bf v}_1])}{r_1^0}
\right]h^2 + O(h^2)
\nonumber\\[1ex]
{\cal C}_0&=&\frac{-c_1^0}{q(r_1^0)^2r_2^0}
\\
&+&\frac{-c_1^0}{2q(r_1^0)^2r_2^0}\left[
\frac{[{\bf n}_q{\bf v}_2]^2}{(r_2^0)^2}+
2\frac{([{\bf n}_q{\bf v}_1][{\bf n}_q{\bf v}_2])}{r_1^0r_2^0}
+3\frac{[{\bf n}_q{\bf v}_1]^2}{(r_1^0)^2}
\right]h^2
\nonumber\\
&+&\frac{1}{2q(r_1^0)^2r_2^0}\left[
\frac{([{\bf n}_q{\dot {\bf v}}_1][{\bf n}_q{\bf v}_2])}{r_2^0}+
2\frac{([{\bf n}_q{\dot {\bf v}}_1][{\bf n}_q{\bf v}_1])}{r_1^0}
\right]h^2 + O(h^2)
\nonumber
\end{eqnarray}

The last expansion we shall need is
\begin{eqnarray}
{\cal D}_b^i&=&k_b^3n_q^i{\cal D}_0
\\
&+&
\frac{h^2}{2qr_1^0r_2^0}\left[
\frac{v_2^i-n_q^i({\bf n}_q{\bf v}_2)}{r_2^0}+
\frac{v_1^i-n_q^i({\bf n}_q{\bf v}_1)}{r_1^0}
\right]
\nonumber\\
&+&
\frac{h^4}{8qr_1^0r_2^0}\left\{
3\frac{[v_2^i-n_q^i({\bf n}_q{\bf v}_2)][{\bf n}_q{\bf
v}_2]^2}{(r_2^0)^3}\right.
\nonumber\\
&+&\left.\frac{2[v_2^i-n_q^i({\bf n}_q{\bf v}_2)]([{\bf n}_q{\bf
v}_1][{\bf n}_q{\bf v}_2])+
[v_1^i-n_q^i({\bf n}_q{\bf v}_1)][{\bf n}_q{\bf
v}_2]^2}{r_1^0(r_2^0)^2}
\right.
\nonumber\\
&+&\left.\frac{2[v_1^i-n_q^i({\bf n}_q{\bf v}_1)]([{\bf n}_q{\bf
v}_1][{\bf n}_q{\bf v}_2])+
[v_2^i-n_q^i({\bf n}_q{\bf v}_2)][{\bf n}_q{\bf
v}_1]^2}{(r_1^0)^2r_2^0}
\right.
\nonumber\\
&+&\left.
3\frac{[v_1^i-n_q^i({\bf n}_q{\bf v}_1)][{\bf n}_q{\bf
v}_1]^2}{(r_1^0)^3}
\right\}+O(h^4).
\nonumber
\end{eqnarray}
By ${\cal D}_0$ we denote the following expansion:
\begin{eqnarray}
{\cal D}_0&=&\frac{1}{qr_1^0r_2^0}
\\
&+&\frac{h^2}{2qr_1^0r_2^0}\left\{
\frac{[{\bf n}_q{\bf v}_1]^2}{(r_1^0)^2}+
\frac{([{\bf n}_q{\bf v}_1][{\bf n}_q{\bf v}_2])}{r_1^0r_2^0}+
\frac{[{\bf n}_q{\bf v}_2]^2}{(r_2^0)^2}
\right\}
\nonumber\\
&+&\frac{h^4}{8qr_1^0r_2^0}\left\{
3\frac{[{\bf n}_q{\bf v}_1]^4}{(r_1^0)^4}+
3\frac{[{\bf n}_q{\bf v}_1]^2([{\bf n}_q{\bf v}_1][{\bf n}_q{\bf
v}_2])}{(r_1^0)^3r_2^0}\right.\nonumber\\
&+&\left.
\frac{2([{\bf n}_q{\bf v}_1][{\bf n}_q{\bf
v}_2])^2+[{\bf n}_q{\bf v}_1]^2[{\bf n}_q{\bf v}_2]^2}{(r_1^0)^2(r_2^0)^2}
\right.\nonumber\\
&+&\left.
3\frac{[{\bf n}_q{\bf v}_2]^2([{\bf n}_q{\bf v}_1][{\bf n}_q{\bf
v}_2])}{r_1^0(r_2^0)^3}+
3\frac{[{\bf n}_q{\bf v}_2]^4}{(r_2^0)^4}
\right\}+O(h^4).
\nonumber
\end{eqnarray}

Our final task will be to compute expression (\ref{tail_i}). When we
differentiate functions ${\cal B}_b^i$, ${\cal C}_b^i$ and ${\cal D}_b^i$
we must keep in mind that derivatives of $h^2$ with respect to $t_a$
do not vanish even if $h^2\to 0$. With a degree of accuracy sufficient
for our purposes we obtain
\begin{eqnarray}
{\cal A}_1^i-\frac{\partial {\cal B}_1^i}{\partial t_1}
-\frac{\partial {\cal C}_1^i}{\partial t_2}+
\frac{\partial^2 {\cal D}_1^i}{\partial t_1\partial t_2}&=&
v_1^i\left({\cal B}_0-\frac{\partial {\cal D}_0}{\partial t_2}
\right)
\\
{\cal A}_2^i-\frac{\partial {\cal B}_2^i}{\partial t_1}
-\frac{\partial {\cal C}_2^i}{\partial t_2}+
\frac{\partial^2 {\cal D}_2^i}{\partial t_1\partial t_2}&=&
v_2^i\left({\cal C}_0-\frac{\partial {\cal D}_0}{\partial t_1}
\right)
\nonumber
\end{eqnarray}
Substituting these relations into eqs.(\ref{b-terms}) returns the integral
(\ref{Tif}) of interference part of the momentum density $T_{int}^{0i}$
over $\varphi$ as the combination of partial derivatives in time
variables.

\subsubsection{Direct particle fields and Lorenz forces}
\renewcommand{\theequation}{\Alph{subsubsection}.\arabic{equation}}
\setcounter{equation}{0}

In classical electrodynamics the four-dimensional delta function of the
square of the interval between points $A$ and $B$ is Green's function of
the wave operator. The delta function ensures that the typical points $A$
and $B$ on the worldlines of point-like charges $a$ and $b$ interact if
and only if they are connectible by a null ray. The interaction is
described by Lorentz force, i.e. there is no self action.

\begin{figure}[h]
\begin{center}
\epsfclipon
\epsfig{file=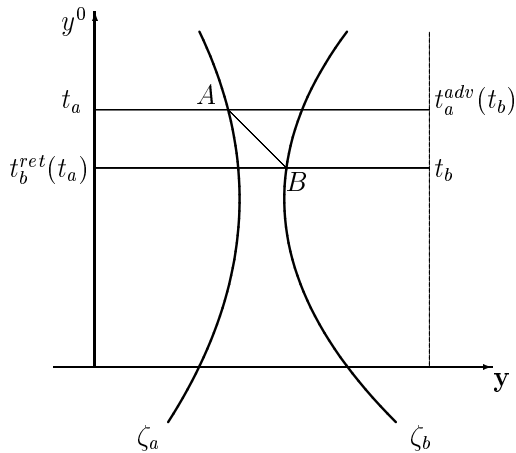,width=7cm}
\end{center}
\caption{\label{A-B}
\small Points $A\in\zeta_a$ and $B\in\zeta_b$ are connectible by a null ray.
They are defined by the pair of instants either $(t_a,t_b^{ret}(t_a))$ or
$(t_a^{adv}(t_b),t_b)$. Functions $t_b^{ret}(t_a)$ and $t_a^{adv}(t_b)$ are
inverses.
}
\end{figure}

The particle $a$ is acted on by the particle $b$ via Lorentz force
$F_{ba}^\alpha=e_aF_{(b)\beta}^\alpha u_a^\beta$ where
$F_{(b)\beta}^\alpha$ is {\it direct particle field} \cite{HN}. By this we
mean electromagnetic field generated by $b$-th particle at point where
$a$-th particle is located. It immediately implies $h=0$ in expressions
(\ref{f}) for the components of electromagnetic fields. Indeed, $h$ is the
radius of the circle $S_a\cap S_b$, i.e. of the intersection of spherical
fronts of outgoing electromagnetic waves generated by charges (see
Figs.\ref{intf}, \ref{k_k}). If we consider the {\it direct particle
field}, the sphere $S_a$ reduces to the point where $a$-th particle is
placed.

To evaluate the retarded field of the 2-nd particle at point
$z_1(t_1)\in\zeta_1$ we put $k_1^0=0$ and $k_2^0=q[t_1,t_2^{ret}(t_1)]$
in (\ref{f}). It implies
\begin{equation}
K_2^0=q,\quad K_2^i=qn_q^i,\quad r_2=q\left[1-V_2\right],\quad
c_2=\gamma_2^{-2} + q{\dot V}_2
\end{equation}
in the expression for $f^{(2)}_{\alpha\beta}$. (It is obvious, that
$f^{(1)}_{\alpha\beta}$ vanishes.) All the quantities are
evaluated at the moments either $t_1$ or $t_2^{ret}(t_1)$,
$V_2:=({\bf n}_q\cdot{\bf v}_2)$ and ${\dot V}_2:=({\bf n}_q\cdot{\bf \dot
v}_2)$.

To find out the advanced field of the 1-st particle at point
$z_2(t_2)\in\zeta_2$, we put $k_2^0=0$ and $k_1^0=-q[t_1^{adv}(t_2),t_2]$
in $f^{(1)}_{\alpha\beta}$ given by eq.(\ref{f}). It means
\begin{equation}
K_1^0=-q,\quad K_1^i=-qn_q^i,\quad r_1=-q\left[1-V_1\right],\quad
c_1=\gamma_1^{-2} - q{\dot V}_1
\end{equation}
where $V_1:=({\bf n}_q\cdot{\bf v}_1)$ and ${\dot V}_1:=({\bf n}_q\cdot{\bf
\dot v}_1)$.

In general, to obtain the retarded/advanced field generated by $a$-th
particle at point where $b$-th particle is located, one should substitute
the quantities
\begin{equation}\label{Krc}
K_a^0=\epsilon q,\quad K_a^i=(-1)^aqn_q^i,\quad
r_a=\epsilon q\left[1-(-1)^a\epsilon V_a\right],\quad
c_a=\gamma_a^{-2} + (-1)^aq{\dot V}_a
\end{equation}
in eq.(\ref{f}). Parameter $\epsilon$ is equal to $+1$ for retarded fields
and $-1$ for advanced ones. Putting eqs.(\ref{Krc}) in (\ref{f}) we
arrive at the following expressions:
\begin{eqnarray}
\!F^{(a)}_{0i}(\epsilon)&=&e_a\left\{\frac{-\varepsilon_a
n_q^i+v_a^i}{q^2\left(1-\varepsilon_aV_a\right)^3}\gamma_a^{-2}
+(-1)^a\frac{-\varepsilon_a
n_q^i+v_a^i}{q\left(1-\varepsilon_aV_a\right)^3}{\dot V}_a
+\epsilon\frac{\dot v_a^i}{q\left(1-\varepsilon_a V_a\right)^2}\right\}
\nonumber\\
\!\!F^{(a)}_{ij}(\epsilon)\!\!\!&=&\!\!\!e_a\left\{
\varepsilon_a\frac{v_a^in_q^j-v_a^jn_q^i}{q^2\left(1-
\varepsilon_aV_a\right)^3}\gamma_a^{-2}+
\epsilon\frac{v_a^in_q^j-v_a^jn_q^i}{q\left(1-
\varepsilon_aV_a\right)^3}{\dot V}_a
+(-1)^a\frac{{\dot v}_a^in_q^j-{\dot v}_a^jn_q^i}{q\left(1-
\varepsilon_aV_a\right)^2}\right\}
\nonumber
\\
\end{eqnarray}
where parameters
\begin{equation}
\varepsilon_a=(-1)^a\epsilon.
\end{equation}

The components of Lorentz force $a$-th charge acting on $b$-th one are
written as follows:
\begin{eqnarray}
\gamma_b^{-1}F^0_{ab}(\epsilon)&=&-e_bF^{(a)}_{0i}(\epsilon)v_b^i
\\
&=&e_be_a\left\{
\frac{\varepsilon_aV_b-({\bf
v}_a\cdot{\bf v}_b)}{q^2\left(1-\varepsilon_aV_a\right)^3}\gamma_a^{-2}
+(-1)^a\frac{\varepsilon_aV_b-({\bf
v}_a\cdot{\bf v}_b)}{q\left(1-\varepsilon_aV_a\right)^3}{\dot V}_a
\right.
\nonumber\\
&-&\left.\epsilon\frac{({\bf\dot
v}_a\cdot{\bf v}_b)}{q\left(1-\varepsilon_aV_a\right)^2}
\right\}
\nonumber\\
\gamma_b^{-1}F^i_{ab}(\epsilon)&=&-e_bF^{(a)}_{0i}(\epsilon) +
e_bF^{(a)}_{ij}(\epsilon)v_b^j
\\
&=&-e_be_a\left\{\varepsilon_an_q^i\left[
\frac{-1+({\bf
v}_a\cdot{\bf v}_b)}{q^2\left(1-\varepsilon_aV_a\right)^3}\gamma_a^{-2}
+(-1)^a\frac{-1+({\bf
v}_a\cdot{\bf v}_b)}{q\left(1-\varepsilon_aV_a\right)^3}{\dot V}_a
\right.\right.
\nonumber\\
&+&\left.\left.\epsilon\frac{({\bf\dot
v}_a\cdot{\bf v}_b)}{q\left(1-\varepsilon_aV_a\right)^2}
\right]\right.\nonumber\\
&+&\left.\frac{1-\varepsilon_aV_b}{1-\varepsilon_aV_a}
\left[
\frac{v_a^i}{q^2\left(1-\varepsilon_aV_a\right)^2}\gamma_a^{-2}
+(-1)^a\frac{v_a^i}{q\left(1-\varepsilon_aV_a\right)^2}{\dot
V}_a
\right.\right.\nonumber\\
&+&\left.\left.\epsilon\frac{\dot{v}_a^i}{q\left(1-\varepsilon_aV_a\right)}
\right]\right\}.\nonumber
\end{eqnarray}
All the quantities labelled by $a$ are referred to the instant
$t_a^\epsilon(t_b)$ while those supplemented with index $b$ are evaluated
at $t_b$.

\subsubsection{Difference of work done by "advanced" and retarded Lorenz
forces}
\renewcommand{\theequation}{\Alph{subsubsection}.\arabic{equation}}
\setcounter{equation}{0}

The retarded, $t_a^{ret}(t_b)$, and "advanced", $t_b^{adv}(t_a)$, instants
arise naturally within the integration procedure developed in Section 5 as
the end points of "inner" integrals (see eqs.(\ref{int1}) and
(\ref{int2})). Typical points $A$ (on the worldline of charge $a$) and $B$
(on the worldline of charge $b$) interact if the line connecting them is a
null ray. It seems, that the interaction can be both forward ($B$ to $A$)
and backward ($A$ to $B$) in time (see Fig.\ref{A-B}). And yet the retarded
causality is not violated. Indeed, we consider the interference of {\it
outgoing} waves present at the observation time $t$. Both the retarded and
"advanced" moments are {\it before} $t$.

In this subsection we compare the work done by retarded Lorentz force due
to charge $b$ on charge $a$
\begin{equation}\label{ret}
\int\limits_{-\infty}^tdt_a \gamma_a^{-1}F_{ba}^\mu[t_a,t_b^{ret}(t_a)]
\end{equation}
and the work done by "advanced" response of charge $a$ on charge $b$
\begin{equation}\label{adv}
\int\limits_{-\infty}^{t_b^{ret}(t)}dt_b\gamma_b^{-1}F_{ab}^\mu
[t_a^{adv}(t_b),t_b].
\end{equation}

The following identity generalises the derivatives of
eqs.(\ref{ret2}), (\ref{adv2}), (\ref{ret1}) and (\ref{adv1}):
\begin{equation}\label{dt}
\frac{dt_a^\epsilon(t_b)}{dt_b}=
\frac{1-\varepsilon_aV_b}{1-\varepsilon_aV_a}.
\end{equation}
Here
\begin{equation}
\varepsilon_a=(-1)^a\epsilon ;
\end{equation}
parameter $\epsilon$ is equal to $+1$ for retarded instants
and $-1$ for advanced ones.
With the help of eq.(\ref{dt}) we obtain the following chain of
identities:
\begin{eqnarray}
\frac{dq[t_a^\epsilon(t_b),t_b]}{dt_b}&=&
(-1)^a\frac{V_b-V_a}{1-\varepsilon_aV_a}
\\\label{dq}
\frac{dn_q^i[t_a^\epsilon(t_b),t_b]}{dt_b}&=&\frac{(-1)^a}{q}
\left[
v_b^i-v_a^i\frac{1-\varepsilon_aV_b}{1-\varepsilon_aV_a}
-n_q^i\frac{V_b-V_a}{1-\varepsilon_aV_a}
\right]\\
\frac{d({\bf n}_q\cdot{\bf v}_b)}{dt_b}&=&\frac{(-1)^a}{q}
\left[
-\gamma_b^{-2}-\left[-1+({\bf v}_1\cdot{\bf v}_2)\right]
\frac{1-\varepsilon_aV_b}{1-\varepsilon_aV_a}
\right.\nonumber\\
&+&\left.\varepsilon_a(V_b-V_a)\frac{1-\varepsilon_aV_b}{1-\varepsilon_aV_a}
\right]+{\dot V}_b
\\
\frac{d({\bf n}_q\cdot{\bf v}_a)}{dt_b}&=&\frac{(-1)^a}{q}
\left[
\gamma_a^{-2}\frac{1-\varepsilon_aV_b}{1-\varepsilon_aV_a}
-1+({\bf v}_1\cdot{\bf v}_2)
+\varepsilon_a(V_b-V_a)\right]\nonumber\\
&+&{\dot V}_a\frac{1-\varepsilon_aV_b}{1-\varepsilon_aV_a}\label{dVa}
\end{eqnarray}

To compare (\ref{ret}) and (\ref{adv}) we change the variables
$[t_a^{adv}(t_b),t_b]\mapsto [t_a,t_b^{ret}(t_a)]$ in "advanced" integral:
\begin{eqnarray}\label{intra}
&&\int\limits_{-\infty}^tdt_a
\gamma_a^{-1}F_{ba}^\mu[t_a,t_b^{ret}(t_a)]
-\int\limits_{-\infty}^{t_b^{ret}(t)}dt_b\gamma_b^{-1}F_{ab}^\mu
[t_a^{adv}(t_b),t_b]
\\
&=&\int\limits_{-\infty}^tdt_a\left[
\gamma_a^{-1}F_{ba}^\mu[t_a,t_b^{ret}(t_a)]
-\left.\frac{1+(-1)^bV_a}{1+(-1)^bV_b}\gamma_b^{-1}F_{ab}^\mu
[t_a^{adv}(t_b),t_b]\right|^{t_a^{adv}(t_b)=t_a}_{t_b=t_b^{ret}(t_a)}\right]
\nonumber
\end{eqnarray}
Using identities (\ref{dq})-(\ref{dVa}) in the integrand of eq.(\ref{intra}),
we derive that it is the total time derivative. In other words, the
difference of "retarded" work (\ref{ret}) and "advanced" one (\ref{adv})
is the integral being a function of the end points only.

\begin{figure}[h]
\begin{center}
\epsfclipon
\epsfig{file=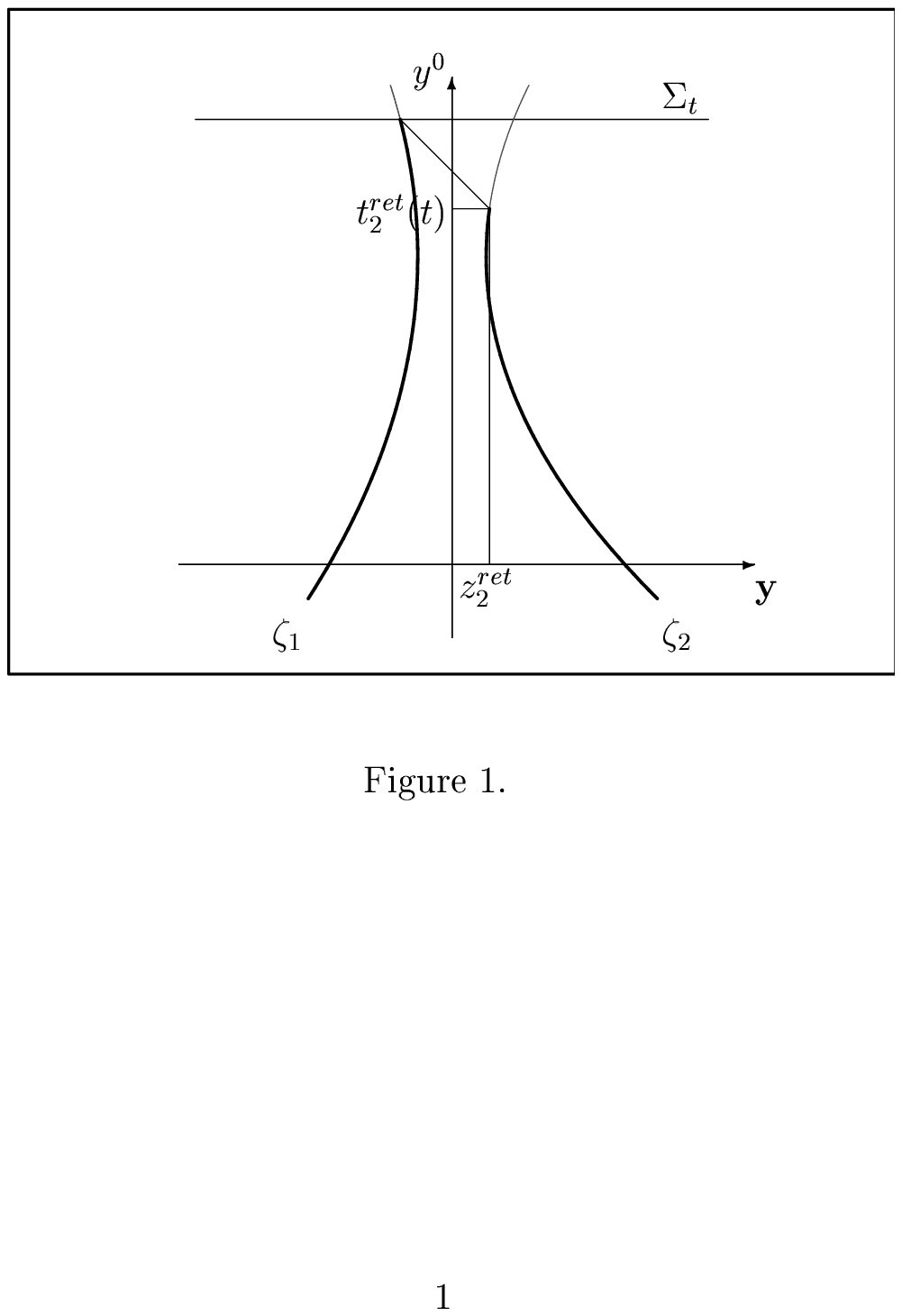,width=7cm}
\end{center}
\caption{\label{work_l}
\small Difference of work done by retarded Lorentz force due to charge 2
on charge 1 and the work done by advanced response of charge 1 is given by
eqs.(\ref{work-l}). All the quantities in the right-hand side
of this equation which are labelled by $1$ are referred to the instant
of observation while those supplemented with index $2$ are evaluated at
$t_2^{ret}(t)$.
}
\end{figure}
\begin{eqnarray}\label{work-l}
&&\int\limits_{-\infty}^tdt_1
\gamma_1^{-1}F_{21}^\mu[t_1,t_2^{ret}(t_1)]
-\int\limits_{-\infty}^{t_2^{ret}(t)}dt_2\gamma_2^{-1}F_{12}^\mu
[t_1^{adv}(t_2),t_2]\\
&=&\left\{
\begin{array}{rcr}
\mu=0&,&-e_1e_2\left[
\frac{\displaystyle -1+({\bf v}_1{\bf v}_2)}{\displaystyle
q[1-V_1][1-V_2]}+
\frac{\displaystyle 1}{\displaystyle q[1-V_1]}+
\frac{\displaystyle 1}{\displaystyle q[1-V_2]}
\right]_{t_1\to -\infty}^{t_1=t}
\\[1em]
\mu=i&,&-e_1e_2\left[
\frac{\displaystyle [-1+({\bf v}_1{\bf v}_2)]n_q^i}{\displaystyle
q[1-V_1][1-V_2]}+
\frac{\displaystyle v_1^i}{\displaystyle q[1-V_1]}+
\frac{\displaystyle v_2^i}{\displaystyle q[1-V_2]}
\right]_{t_1\to -\infty}^{t_1=t}
\end{array}
\right.\nonumber
\end{eqnarray}

\begin{figure}[h]
\begin{center}
\epsfclipon
\epsfig{file=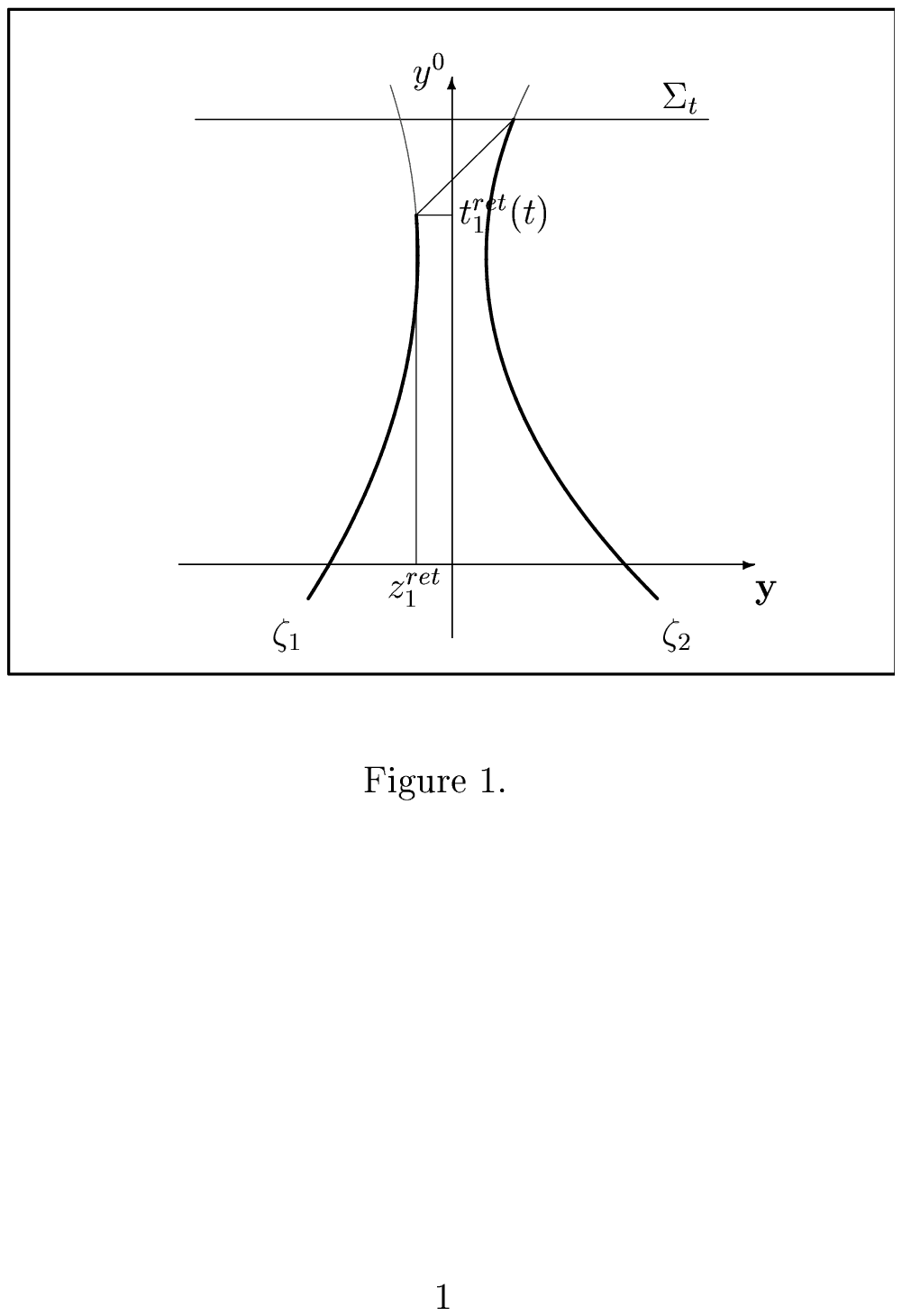,width=7cm}
\end{center}
\caption{\label{work_r}
\small Difference of work done by retarded Lorentz force due to charge 1
on charge 2 and the work done by advanced response of charge 2 is defined
by eqs.(\ref{work-r}). All the quantities in the right-hand side
of this equation which are labelled by $2$ are referred to the instant
of observation while those supplemented with index $1$ are evaluated at
$t_1^{ret}(t)$.
}
\end{figure}
\begin{eqnarray}\label{work-r}
&&\int\limits_{-\infty}^tdt_2
\gamma_2^{-1}F_{12}^\mu[t_1^{ret}(t_2),t_2]
-\int\limits_{-\infty}^{t_1^{ret}(t)}dt_1\gamma_1^{-1}F_{21}^\mu
[t_1,t_2^{adv}(t_1)]\\
&=&\left\{
\begin{array}{rcr}
\mu=0&,&e_1e_2\left[-
\frac{\displaystyle -1+({\bf v}_1{\bf v}_2)}{\displaystyle
q[1+V_1][1+V_2]}-
\frac{\displaystyle 1}{\displaystyle q[1+V_1]}-
\frac{\displaystyle 1}{\displaystyle q[1+V_2]}
\right]_{t_2\to -\infty}^{t_2=t}
\\[1em]
\mu=i&,&e_1e_2\left[
\frac{\displaystyle [-1+({\bf v}_1{\bf v}_2)]n_q^i}{\displaystyle
q[1+V_1][1+V_2]}-
\frac{\displaystyle v_1^i}{\displaystyle q[1+V_1]}-
\frac{\displaystyle v_2^i}{\displaystyle q[1+V_2]}
\right]_{t_2\to -\infty}^{t_2=t}
\end{array}
\right.\nonumber
\end{eqnarray}

It is convenient to rewrite the results (\ref{work-l}) and (\ref{work-r})
in a manifestly covariant fashion:
\begin{eqnarray}\label{d_wrk}
&&\int\limits_{-\infty}^tdt_a
\gamma_a^{-1}F_{ba}^\mu[t_a,t_b^{ret}(t_a)]
-\int\limits_{-\infty}^{t_b^{ret}(t)}dt_b\gamma_b^{-1}F_{ab}^\mu
[t_a^{adv}(t_b),t_b]
\nonumber\\
&=&(-1)^ae_1e_2\left[
\frac{(u_1\cdot u_2)n_q^\mu}{q(n_q\cdot u_1)(n_q\cdot u_2)}
-\frac{u_1^\mu}{q(n_q\cdot u_1)}-\frac{u_2^\mu}{q(n_q\cdot u_2)}
\right]_{t_a\to -\infty}^{t_a=t}
\end{eqnarray}
Symbols $u_a^\mu, a=1,2,$ denotes the (normalized) four-velocity vector
$(\gamma_a^{-1},\gamma_a^{-1}v_a^i)$. If the 2-nd particle moves in the
retarded field of the 1-st one while the 1-st particle moves in the
"advanced" field of the 2-nd one, then $n_q^\mu=(1,n_q^i)$. Four-products
of this null vector with four velocities are as follows:
\begin{equation}
(n_q\cdot u_a)=-\gamma_a^{-1}\left[1-V_a\right].
\end{equation}
If one interchanges the words "first particle" and "second particle" in
the above sentences, $n_q^\mu=(-1,n_q^i)$ and we have
\begin{equation}
(n_q\cdot u_a)=\gamma_a^{-1}\left[1+V_a\right] .
\end{equation}
in eq.(\ref{d_wrk}).

\subsubsection{Time integration of $T_{\rm int}^{0\mu}$}
\renewcommand{\theequation}{\Alph{subsubsection}.\arabic{equation}}
\setcounter{equation}{0}

In this paper we integrate the interference part of energy-momentum tensor
density of two point-like charged particles over three-dimensional
hyperplane $\Sigma_t=\{y\in{\mathbb M}_4: y^0=t\}$. An integration
hypersurface is a surface of constant value of the obsevation time
parameter. Besides $t$, the set of curvilinear coordinates includes the
"individual" retarded times $t_1$ and $t_2$, associated with the
particles' worldlines, and the angle variable $\varphi$. The integration
over $\varphi$ is performed in Appendix A and Appendix B. The crucial
issue is that the resulting expressions are the sum of partial
derivatives in individual times (see eqs.(\ref{fiT00}) and
combination of (\ref{b-terms}) and (\ref{v-terms})). It allows us
to perform the
integration over one of the time parameters, either $t_1$ or
$t_2$. "Retarded" shifts in arguments of particles' individual
characteristics such us coordinates, velocities etc. appear on this stage
as well as "advanced" ones.

The first double integral involved in the rules either (\ref{int1}) or
(\ref{int2}) defines the integration over "causal" region which is
pictured in Figs.\ref{ret_adv} and \ref{ra_cr}, while the second one deals
with "acausal" region (see Figs.\ref{n-caus} and \ref{nc-cr}). The
integration of "causal" type can be handled via the relations
(\ref{dt})-(\ref{dVa}). Their counterparts for "acausal" region look as
follows:
\begin{eqnarray}
\frac{dt_1'(t,t_2)}{dt_2}&=&
-\frac{1-V_2}{1+V_1}\\\label{t1'}
\frac{dq[t_1'(t,t_2),t_2]}{dt_2}&=&
-\frac{V_1+V_2}{1+V_1}
\\
\frac{dn_q^i[t_1'(t,t_2),t_2]}{dt_2}&=&\frac{1}{q}
\left[
-v_2^i-v_1^i\frac{1-V_2}{1+V_1}
+n_q^i\frac{V_1+V_2}{1+V_1}
\right]\\
\frac{d({\bf n}_q\cdot{\bf v}_1)}{dt_2}&=&\frac{1}{q}
\left[
\gamma_1^{-2}\frac{1-V_2}{1+V_1}
-\left[1+({\bf v}_1\cdot{\bf v}_2)\right]+
V_1+V_2
\right]\nonumber\\
&-&{\dot V}_1\frac{1-V_2}{1+V_1}
\\
\frac{d({\bf n}_q\cdot{\bf v}_2)}{dt_2}&=&\frac{1}{q}
\left[
\gamma_2^{-2}-\left[1+({\bf v}_1\cdot{\bf v}_2)\right]
\frac{1-V_2}{1+V_1}
\right.\nonumber\\
&-&\left.(V_1+V_2)\frac{1-V_2}{1+V_1}
\right]+{\dot V}_2
\end{eqnarray}
where $V_a:=({\bf n}_q\cdot{\bf v}_a)$ and ${\dot V}_a:=({\bf n}_q\cdot{\bf
{\dot v}}_a)$.

The way of integration where all the mixed derivatives are written as
$\partial /\partial t_1[\partial G_0/\partial t_2]$ results the
expressions placed in the left columnes of Tables. We apply the rules
(\ref{rt1}), (\ref{ad1}) and (\ref{ac2}) for the 1-st, 2-nd and 3-rd line,
respectively.

If one changes the order of differentiations they obtain the
expressions in the right columnes of Tables. For the 1-st and 2-nd lines
time integration rules are as follows:
\begin{equation} \label{2nd}
\int\limits_{-\infty}^{t_2^{ret}(t)}d
t_2\left[G_1-\frac{1-V_2}{1-V_1}G_2\right]^{t_1=t_1^{adv}(t_2)}\quad
{\rm (a)},\!\!
\int\limits_{-\infty}^td
t_2\left[-G_1+\frac{1+V_2}{1+V_1}G_2\right]_{t_1=t_1^{ret}(t_2)} \quad
{\rm (b)}.\!\!
\end{equation}
Acausal region is integrated according to the rule (\ref{ac2}) (3-rd line
of the right column).
\newpage

{\bf Table 1.} {\small Integral $\int_0^{2\pi}\sqrt{-g}T^{00}_{int}$ has
the form
of $\partial G_1/\partial t_1+\partial G_2/\partial t_2+\partial^2
G_0/\partial t_1\partial t_2$. Integration over time results the
expressions in the left column (if mixed derivative is coupled with
$\partial G_1/\partial t_1$) or in the right column (if $\partial^2
G_0/\partial t_1\partial t_2$ is added to $\partial G_2/\partial t_2$).
Integration over "acausal" region gives the functions of the end points
only (see third line).}

\begin{tabular}{|c|cc|}
\hline
&&\\[-1.5ex]
$\frac{\displaystyle \partial}{\displaystyle \partial
t_1}\left[\frac{\displaystyle \partial G_0}{\displaystyle
\partial t_2}\right]$&
$\frac{\displaystyle \partial}{\displaystyle \partial
t_2}\left[\frac{\displaystyle \partial G_0}{\displaystyle \partial
t_1}\right]$&\\
&&\\[-1.5ex]
\hline
\hline
&&\\[-1.5ex]
(\ref{rt1})$\quad
-\int_{-\infty}^tdt_1\gamma_1^{-1}F^0_{21}[t_1,t_2^{ret}(t_1)]
\phantom{\frac{\displaystyle 1}{\displaystyle 1}}$
&(\ref{2nd}a)
$-\int_{-\infty}^{t_2^{ret}(t)}dt_2\gamma_2^{-1}F^0_{12}[t_1^{adv}(t_2),
t_2]$&\\[1ex]
$+e_1e_2\left[\frac{\displaystyle 1}{\displaystyle
2k_2^0}\frac{\displaystyle 1+V_2}{\displaystyle 1-V_2}
-\frac{\displaystyle 1}{\displaystyle q[1-V_2]}
\right]_{t_1\to -\infty}^{t_1=t}$&
$+e_1e_2\left[\frac{\displaystyle 1}{\displaystyle 2k_1^0}
\frac{\displaystyle 1+V_1}{\displaystyle 1-V_1}
+\frac{\displaystyle 1}{\displaystyle q[1-V_1]}
\right]_{t_2\to
-\infty}^{t_2\to t_2^{ret}(t)}$&
\\
&&\\[-1.5ex]
\hline
&&\\[-1.5ex]
(\ref{ad1})$\quad -\int_{-\infty}^{t_1^{ret}(t)}dt_1\gamma_1^{-1}
F^0_{21}[t_1,t_2^{adv}(t_1)]$&(\ref{2nd}b)
$-\int_{-\infty}^tdt_2\gamma_2^{-1}F^0_{12}[t_1^{ret}(t_2),t_2]
\phantom{\frac{\displaystyle 1}{\displaystyle 1}}$&
\\[1ex]
$+e_1e_2\left[\frac{\displaystyle 1}{\displaystyle
2k_2^0}\frac{\displaystyle 1-V_2}{\displaystyle 1+V_2}
+\frac{\displaystyle 1}{\displaystyle q[1+V_2]}
\right]_{t_1\to -\infty}^{t_1\to t_1^{ret}(t)}$&
$+e_1e_2\left[\frac{\displaystyle 1}{\displaystyle 2k_1^0}
\frac{\displaystyle 1-V_1}{\displaystyle 1+V_1}
-\frac{\displaystyle 1}{\displaystyle q[1+V_1]}
\right]_{t_2\to
-\infty}^{t_2=t}$&
\\
&&\\[-1.5ex]
\hline
&&\\[-1.5ex]
(\ref{ac2})$\qquad -\left.\frac{\displaystyle e_1e_2}{\displaystyle
2k_2^0}
\right|_{t_2=t_2^{ret}(t)}^{t_2\to t}\qquad$&(\ref{ac2})
$\qquad\left.\frac{\displaystyle e_1e_2}{\displaystyle 2k_1^0}
\right|_{t_2\to t_2^{ret}(t)}^{t_2=t}$&
\\
\hline
\end{tabular}

{\bf Table 2.} {\small Integral $\int_0^{2\pi}\sqrt{-g}T^{0i}_{int}$
becomes the combination of partial derivatives in time variables.
Structure of this Table is analogous to the structure of Table 1.}

\bigskip

\begin{tabular}{|c|c|}
\hline
&\\[-1.5ex]
$\frac{\displaystyle \partial}{\displaystyle \partial
t_1}\left[\frac{\displaystyle \partial\Lambda}{\displaystyle
\partial t_2}\right]$&
$\frac{\displaystyle \partial}{\displaystyle \partial
t_2}\left[\frac{\displaystyle \partial\Lambda}{\displaystyle \partial
t_1}\right]$\\
&\\[-1.5ex]
\hline
\hline
&\\[-1.5ex]
$-\int_{-\infty}^tdt_1\gamma_1^{-1}F^i_{21}[t_1,t_2^{ret}(t_1)]
\phantom{\frac{\displaystyle 1}{\displaystyle 1}}$
&
$-\int_{-\infty}^{t_2^{ret}(t)}dt_2\gamma_2^{-1}F^i_{12}[t_1^{adv}(t_2),
t_2]$\\[1ex]
$+e_1e_2\left[\frac{\displaystyle n_q^i+v_2^i}{\displaystyle
2k_2^0[1-V_2]}
-\frac{\displaystyle v_2^i}{\displaystyle q[1-V_2]}
\right]_{t_1\to -\infty}^{t_1=t}$&
$+e_1e_2\left[\frac{\displaystyle n_q^i+v_1^i}{\displaystyle
2k_1^0[1-V_1]}
+\frac{\displaystyle v_1^i}{\displaystyle q[1-V_1]}
\right]_{t_2\to
-\infty}^{t_2\to t_2^{ret}(t)}$
\\
&\\[-1.5ex]
\hline
&\\[-1.5ex]
$-\int_{-\infty}^{t_1^{ret}(t)}dt_1\gamma_1^{-1}
F^i_{21}[t_1,t_2^{adv}(t_1)]$&
$-\int_{-\infty}^tdt_2\gamma_2^{-1}F^i_{12}[t_1^{ret}(t_2),t_2]
\phantom{\frac{\displaystyle 1}{\displaystyle 1}}$
\\[1ex]
$+e_1e_2\left[\frac{\displaystyle -n_q^i+v_2^i}{\displaystyle
2k_2^0[1+V_2]}
+\frac{\displaystyle v_2^i}{\displaystyle q[1+V_2]}
\right]_{t_1\to -\infty}^{t_1\to t_1^{ret}(t)}$&
$+e_1e_2\left[\frac{\displaystyle -n_q^i+v_1^i}{\displaystyle
2k_1^0[1+V_1]}
-\frac{\displaystyle v_1^i}{\displaystyle q[1+V_1]}
\right]_{t_2\to
-\infty}^{t_2=t}$
\\
&\\[-1.5ex]
\hline
&\\[-1.5ex]
$e_1e_2\left.\frac{\displaystyle n_q^i-v_2^i}{\displaystyle 2k_2^0[1-V_2]}
\right|_{t_2=t_2^{ret}(t)}^{t_2\to t}$&
$e_1e_2\left.\frac{\displaystyle n_q^i+v_1^i}{\displaystyle 2k_1^0[1+V_1]}
\right|_{t_2\to t_2^{ret}(t)}^{t_2=t}$
\\[3ex]
\hline
\end{tabular}

Taking into account the relationship (\ref{d_wrk}) between work of the
"advanced" Lorentz force and the work of the "retarded" one we remove all
the "advanced" integrals from these Tables. The final expressions are
written in Table 3.

\newpage

{\bf Table 3.} {\small The expressions which are placed above double line
concern with integration of energy density $T^{00}_{int}$ while ones
below double line result from integration of $T^{0i}_{int}$.}

\begin{tabular}{|c|c|}
\hline
&\\[-1.7ex]
$\frac{\displaystyle \partial}{\displaystyle \partial
t_1}\left[\frac{\displaystyle \partial\Lambda}{\displaystyle
\partial t_2}\right]$&
$\frac{\displaystyle \partial}{\displaystyle \partial
t_2}\left[\frac{\displaystyle \partial\Lambda}{\displaystyle \partial
t_1}\right]$\\
&\\[-1.7ex]
\hline
\hline
&\\[-1.7ex]
$-\int_{-\infty}^tdt_1\gamma_1^{-1}F^0_{21}[t_1,t_2^{ret}(t_1)]$
&
$-\int_{-\infty}^tdt_1\gamma_1^{-1}F^0_{21}[t_1,t_2^{ret}(t_1)]$
\\[1ex]
$+e_1e_2\left[\frac{\displaystyle 1}{\displaystyle 2k_2^0}\frac{\displaystyle
1+V_2}{\displaystyle 1-V_2}
-\frac{\displaystyle 1}{\displaystyle q[1-V_2]}
\right]_{t_1\to -\infty}^{t_1=t}$&
$+e_1e_2\left[\frac{\displaystyle 1}{\displaystyle 2k_1^0}
\frac{\displaystyle 1+V_1}{\displaystyle 1-V_1}
-\frac{\displaystyle 1}{\displaystyle q[1-V_2]}\right.$
\\
&
$\left.-\frac{\displaystyle -1+({\bf v}_1{\bf v}_2)}{\displaystyle
q[1-V_1][1-V_2]}
\right]_{t_1\to
-\infty}^{t_1\to t}$\\
&\\[-1.7ex]
\hline
&\\[-1.7ex]
$-\int_{-\infty}^tdt_2\gamma_2^{-1}F^0_{12}[t_1^{ret}(t_2),t_2]$&
$-\int_{-\infty}^tdt_2\gamma_2^{-1}F^0_{12}[t_1^{ret}(t_2),t_2]$
\\[1ex]
$+e_1e_2\left[\frac{\displaystyle 1}{\displaystyle 2k_2^0}\frac{\displaystyle
1-V_2}{\displaystyle 1+V_2}
-\frac{\displaystyle 1}{\displaystyle q[1+V_1]}\right.$
&
$+e_1e_2\left[\frac{\displaystyle 1}{\displaystyle 2k_1^0}
\frac{\displaystyle 1-V_1}{\displaystyle 1+V_1}
-\frac{\displaystyle 1}{\displaystyle q[1+V_1]}
\right]_{t_2\to
-\infty}^{t_2=t}$
\\
$\left.-\frac{\displaystyle -1+({\bf v}_1{\bf v}_2)}{\displaystyle
q[1+V_1][1+V_2]}
\right]_{t_2\to -\infty}^{t_2\to t}$&\\
&\\[-1.7ex]
\hline
&\\[-1.7ex]
$-\left.\frac{\displaystyle e_1e_2}{\displaystyle 2k_2^0}
\right|_{t_2=t_2^{ret}(t)}^{t_2\to t}$&
$\left.\frac{\displaystyle e_1e_2}{\displaystyle 2k_1^0}
\right|_{t_2\to t_2^{ret}(t)}^{t_2=t}$
\\
&\\[-1.7ex]
\hline\hline
&\\[-1.7ex]
$-\int_{-\infty}^tdt_1\gamma_1^{-1}F^i_{21}[t_1,t_2^{ret}(t_1)]$
&
$-\int_{-\infty}^tdt_1\gamma_1^{-1}F^i_{21}[t_1,t_2^{ret}(t_1)]$
\\[1ex]
$+e_1e_2\left[\frac{\displaystyle n_q^i+v_2^i}{\displaystyle
2k_2^0[1-V_2]}
-\frac{\displaystyle v_2^i}{\displaystyle q[1-V_2]}
\right]_{t_1\to -\infty}^{t_1=t}$&
$+e_1e_2\left[\frac{\displaystyle n_q^i+v_1^i}{\displaystyle
2k_1^0[1-V_1]}
-\frac{\displaystyle v_2^i}{\displaystyle q[1-V_2]}
\right.$
\\
&
$\left.
-\frac{\displaystyle\left[-1+({\bf v}_1{\bf
v}_2)\right]n_q^i}{\displaystyle q[1-V_1][1-V_2]}
\right]_{t_1\to -\infty}^{t_1\to t}$
\\
&\\[-1.7ex]
\hline
&\\[-1.7ex]
$-\int_{-\infty}^tdt_2\gamma_2^{-1}F^i_{12}[t_1^{ret}(t_2),t_2]$
&
$-\int_{-\infty}^tdt_2\gamma_2^{-1}F^i_{12}[t_1^{ret}(t_2),t_2]$
\\[1ex]
$+e_1e_2\left[\frac{\displaystyle -n_q^i+v_2^i}{\displaystyle
2k_2^0[1+V_2]}
-\frac{\displaystyle v_1^i}{\displaystyle q[1+V_1]}\right.$
&$+e_1e_2\left[\frac{\displaystyle -n_q^i+v_1^i}{\displaystyle
2k_1^0[1+V_1]}
-\frac{\displaystyle v_1^i}{\displaystyle q[1+V_1]}
\right]_{t_2\to
-\infty}^{t_2=t}$
\\
$\left.
+\frac{\displaystyle\left[-1+({\bf v}_1{\bf
v}_2)\right]n_q^i}{\displaystyle q[1+V_1][1+V_2]}
\right]_{t_2\to -\infty}^{t_2\to t}$&
\\
&\\[-1.7ex]
\hline
&\\[-1.7ex]
$e_1e_2\left.\frac{\displaystyle n_q^i-v_2^i}{\displaystyle 2k_2^0[1-V_2]}
\right|_{t_2=t_2^{ret}(t)}^{t_2\to t}$&
$e_1e_2\left.\frac{\displaystyle n_q^i+v_1^i}{\displaystyle 2k_1^0[1+V_1]}
\right|_{t_2\to t_2^{ret}(t)}^{t_2=t}$
\\[3ex]
\hline
\end{tabular}

\end{document}